\pdfoutput=1
\documentclass[a4paper,twocolumn]{article}
\usepackage{abstract}
\usepackage[T1]{fontenc}
\usepackage{booktabs}
\usepackage{amsmath,amssymb}
\usepackage{xfrac}
\allowdisplaybreaks[3]
\newcommand{\vect}[1]{\boldsymbol{#1}}

\usepackage{graphicx}
\usepackage[font=small,labelfont=bf]{caption}
\usepackage{subfig}
\usepackage{tikz}
\usetikzlibrary{arrows}
\usetikzlibrary{positioning}
\usetikzlibrary{calc}
\usepackage{pgfplots}
\usepgflibrary{fpu}
\usepackage{pgfplotstable}
\usepackage{multirow}
\usepackage{rotating}

\newlength{\picwidth}
\pgfplotsset{compat=1.5.1}

\usepackage{dblfloatfix}
\renewcommand{\topfraction}{0.85}
\renewcommand{\textfraction}{0.1}
\renewcommand{\floatpagefraction}{0.75}

\newcommand\T{\rule{0pt}{2.6ex}}
\newcommand\B{\rule[-1.2ex]{0pt}{0pt}}

    \renewcommand{\topfraction}{0.9}	
    \renewcommand{\bottomfraction}{0.8}	
    \renewcommand{\dbltopfraction}{0.9}	
    \renewcommand{\textfraction}{0.07}	
    \renewcommand{\floatpagefraction}{0.7}	
    \renewcommand{\dblfloatpagefraction}{0.7}	

\usepackage[backend=bibtex,style=authoryear,sorting=nty,natbib]{biblatex}
\bibliography{refs.bib}
\newcommand{\sref}[1]{\S\ref{#1}}
\newcommand{\tref}[1]{\tablename~\ref{#1}}
\newcommand{\fref}[1]{\figurename~\ref{#1}}

\newcommand{\abs}{\mathrm{abs}}
\newcommand{\rad}{\,\mathrm{rad}}
\newcommand{\m}{\,\mathrm{m}}
\newcommand{\km}{\,\mathrm{km}}
\newcommand{\nm}{\,\mathrm{nm}}
\newcommand{\cm}{\,\mathrm{cm}}
\newcommand{\um}{\,\mathrm{\mu m}}
\newcommand{\hz}{\,\mathrm{Hz}}
\newcommand{\px}{\,\mathrm{px}}

\def\rand#1#2{\ensuremath{\mathrm{uniform\lbrack#1,#2\rbrack}}}

\def\spose#1{\hbox to 0pt{#1\hss}}
\newcommand{\lta}{\mathrel{\spose{\lower 3pt\hbox{$\mathchar"218$}}
    \raise 2.0pt\hbox{$\mathchar"13C$}}}
\newcommand{\gta}{\mathrel{\spose{\lower 3pt\hbox{$\mathchar"218$}}
    \raise 2.0pt\hbox{$\mathchar"13E$}}}

\let\lesssim=\lta
\let\gtrsim=\gta

\newcommand{\phh}{0.8}
\newcommand{\dza}{1}
\newcommand{\dzb}{2.5}
\newcommand{\tlhl}{3}

\tikzstyle{plane} = [rectangle, draw, fill=blue!20, 
    text width=0.5\columnwidth, text centered, rounded corners, minimum height=4em]
\tikzstyle{stage} = [rectangle, draw, fill=black!20, 
    text width=0.5\columnwidth, text centered, rounded corners, minimum height=4em]
\tikzstyle{start} = [rectangle, draw, fill=red!20, 
    text centered, rounded corners, minimum height=2em]
\tikzstyle{finish} = [rectangle, draw, fill=green!20, 
    text centered, rounded corners, minimum height=2em]
\tikzstyle{line} = [draw, rounded corners, ->, >=stealth]
\tikzstyle{plabel} = [rectangle, text width=7em]
\tikzstyle{plabel1} = [rectangle, text width=8em]
\tikzstyle{plabel2} = [rectangle, text width=4.5em]
\tikzstyle{quartile} = [dotted,mark=none]
\tikzstyle{minmax} = [dotted,mark=none]
\tikzstyle{med} = [solid,mark=none]
\tikzstyle{pl} = [rectangle, draw=none, fill=blue!20, inner sep=0,
    text width=0.5\columnwidth, text centered, rounded corners, minimum height=0.7\columnwidth]
\tikzstyle{pllabel}=[anchor=north,text centered,minimum height=2em]
\tikzstyle{inp}=[text width=1.5ex,inner sep=0,draw, shape=circle]
\tikzstyle{outp}=[text width=1.5ex,fill=black,inner sep=0, shape=circle]

\begin{document}
\nocite{*}
\title{Wavefront Phase Retrieval with Non-linear Curvature Sensors}
\author{P.~L.~Aisher, J.~Crass, C.~Mackay, Inst of Astronomy, University of Cambridge}

\twocolumn[
\maketitle
\begin{onecolabstract}
Increasing interest in astronomical applications of non-linear curvature wavefront sensors for turbulence detection and correction makes it important to understand how best to handle the data they produce, particularly at low light levels. Algorithms for wavefront phase-retrieval from a four-plane curvature wavefront sensor are developed and compared, with a view to their use for low order phase compensation in instruments combining adaptive optics and Lucky Imaging. The convergence speed and quality of iterative algorithms is compared to their step-size and techniques for phase retrieval at low photon counts are explored. 

Computer simulations show that at low light levels, preprocessing by convolution of the measured signal with a gaussian function can reduce by an order of magnitude the photon flux required for accurate phase retrieval of low-order errors. This facilitates wavefront correction on large telescopes with very faint reference stars.
\medskip
\end{onecolabstract}
]

\section{Introduction}
\label{sec:intro}

Adaptive optics (AO) systems, following their proposal by \citet{Babcock},  have been used successfully on large ground-based telescopes to correct for the effects of atmospheric turbulence on  incoming wavefronts, particularly in the infrared \autocite{Beckers}. The majority of these systems use a Shack-Hartmann wavefront sensor \autocite{Shack71,Hartman00} to derive the wavefront  phase error  data for the AO system, with typically $8 \leq N \leq 64$ subapertures across the telescope diameter.

Lucky Imaging (LI), a term coined by \citet{Fried78}, is a technique to provide diffraction-limited images from ground-based telescopes. Many short-exposure images are taken, and a subset are selected on the basis of the sharpness of a reference star in the field of view.

In Lucky Imaging applications, which are being used increasingly, removal of as much aberration as possible is desirable, regardless of the scale-length of phase distortions.  On telescopes larger than 2.5m, Lucky Imaging will only work in the visible when combined with some additional degree of low-order phase correction such as may be provided by an adaptive optics system.  In such a system, when using Shack-Hartmann wavefront sensors (SHWFS) the number of lenslets is selected based upon the degree of correction required.  With highly aberrated wavefronts, many more lenslets are needed than for less aberrated wavefronts. The chosen number of lenslets is very difficult to change once the system has been built. Because of the fixed number of sub-apertures, and because of their fixed focal length, SHWFS systems cannot simultaneously correct for all length-scales of aberration equally well.  Curvature wavefront sensors (CWFS) have been shown to outperform SHWFS systems in some circumstances, especially in correcting low- and high-order aberrations simultaneously using a non-linear curvature wavefront sensor (nlCWFS) \autocite{Guyon10}. 

The curvature wavefront methods are extensions of a broad class of phase diversity wavefront sensing strategies \autocite{Gonsalves82}. In essence, these methods measure the intensity of an aberrated wavefront as it propagates through an optical system. Two or more of these intensity measurements can then be used to derive the true phase of the wavefront entering the system. nlCWFS systems measure the wavefront intensity in four planes, comprised of two equidistant pairs of planes, parallel to, and at different distances from the pupil plane. In the pupil plane the light intensity is typically assumed to be uniform.   On either side of the pupil the light breaks up into a pattern of diffraction limited speckles.  As light propagates through the  pupil, a region which  changes from bright to dark implies a divergent wavefront while one that changes from dark to bright corresponds to a convergent wavefront. An iterative reconstruction algorithm is then used to reconstruct the pupil phase. 

Two planes used in a traditional CWFS are sufficient for this technique to work, but using four has a potential advantage as it allows the non-linear propagation of light through the pupil plane to be properly modelled. High-order phase distortions will introduce intensity changes on a shorter length-scale than low-order phase distortions. Using a pair of planes close to the pupil plane, and a pair of planes far from the pupil plane, different length-scales of phase distortion can be recorded and corrected \autocite{Guyon10}.

The earliest algorithm for phase-retrieval from intensity measurements in different planes, the Error-Reduction (ER) algorithm \autocite{GerchSax72}, worked with data both in the image and Fourier domains. For a high signal-to-noise input signal, the RMS error in the phase estimate was shown mathematically to decrease at every iteration, although the reduction for each step quickly became very small. A series of alternative algorithms have been proposed, all abstractions from the ER algorithm \autocite{Fienup82}, which viewed a portion of the iterative process as a linear function, and applied gradient-search techniques. These algorithms have much faster convergence, but with that faster convergence comes the potential for instability. 

The initial development of the case of four planes nlCWFS used an ER algorithm approach to recover phase from intensity measurements in post processing \autocite{Guyon10}. To facilitate the use of the method in real-time, the stagnation of convergence of the ER method must be overcome with one possible method being to use the techniques proposed by Fienup. 

The course followed here has been to apply a modification to the Fienup algorithms to the case of four images planes offering an improvement in the convergence rate of the wavefront fitting, particularly at low signal-to-noise.  This is especially important as the ability of a nlCWFS to work effectively at the lowest signal levels will enable much fainter reference stars to be used in the adaptive optics correction system.  This is important if the fraction of the sky accessible for AO assisted science observations is to be maximised.  Presently, a relatively bright reference star is required by AO wavefront sensors, limiting their application using natural guide stars to much less than 1\% of the sky, even when non-linear phase retrieval strategies are used \autocite{ClareLane2004}.

Although it is  important to consider the computational feasibility of each approach to wavefront sensing, it is worth bearing in mind that the limit of what is possible with computers is a moving target. Order of magnitude estimations for what is currently possible are useful, but precise values based on today's technology will quickly become obsolete. Some techniques presented in this paper are probably beyond the realms of today's silicon, but nevertheless offer performance enhancements. These might currently only be interesting side-notes, but over the space of the next decade, it is likely that Moore's `Law' will continue its steady progress, changing what can and cannot be considered as viable options.

Most approaches, CWFS included, work very well with many photons, but in Lucky Imaging applications, we can expect to encounter situations with relatively few. Reconstruction algorithms struggle here as a pixel with zero intensity can carry no phase information. This means that the algorithm is constantly throwing away phase data by clamping intensities back to zero. It should be emphasised, however, that total phase correction is unnecessary. Lucky Imaging has been shown to work well on $2.5\m$ telescopes, where  there may be seven or eight turbulent cells across the diameter ($D\sim8r_0$). Obtaining this level of error is all that is required therefore --- the statistics will do the rest.  It is also important to remember that for Lucky Imaging we only wish to improve the images so that a good percentage are of adequate quality.  We are not at all trying to achieve a perfect compensation.

In \sref{sec:instr} instrumental configurations and requirements are described. In \sref{sec:alg} the workings of the algorithm and some possible variations to it, including a parallelised implementation, are explained. In \sref{sec:lowlight}, strategies for dealing with low photon counts are presented and quantitatively compared. An experimental comparison of different iterative algorithms is presented in \sref{sec:exp}, and conclusions are drawn in \sref{sec:discuss}.


\section{Instrument}
\label{sec:instr}

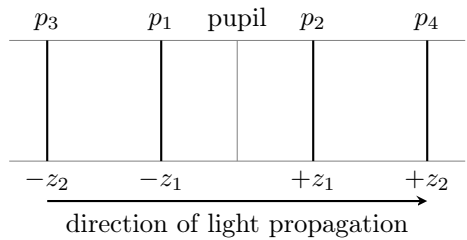
\begin{figure}[h]
\caption{Geometry of the measurement planes}
\centering
\begin{tikzpicture}[>=stealth]
\label{fig:planediagram}
\begin{scope}[help lines]
  \draw (-\tlhl,-\phh) -- (\tlhl,-\phh);
  \draw (-\tlhl,+\phh) -- (\tlhl,+\phh);
  \draw (0,-\phh) -- (0,+\phh);        
\end{scope}
  \node[above] at (0,+\phh) {pupil};
\begin{scope}[thick]
  \draw[thick] (-\dza,-\phh) -- (-\dza,+\phh);        
  \node[below] at (-\dza,-\phh) {$-z_1$};
  \node[above] at (-\dza,+\phh) {$p_1$};
  \draw[thick] (+\dza,-\phh) -- (+\dza,+\phh);      
  \node[below] at (+\dza,-\phh) {$+z_1$};
  \node[above] at (+\dza,+\phh) {$p_2$};
  \draw[thick] (-\dzb,-\phh) -- (-\dzb,+\phh);      
  \node[below] at (-\dzb,-\phh) {$-z_2$};
  \node[above] at (-\dzb,+\phh) {$p_3$};    
  \draw[thick] (+\dzb,-\phh) -- (+\dzb,+\phh);      
  \node[below] at (+\dzb,-\phh) {$+z_2$};
  \node[above] at (+\dzb,+\phh) {$p_4$};
  \draw[->,thick] (-\dzb,-\phh cm - 3.5 ex) -- (+\dzb,-\phh cm - 3.5 ex)
    node[midway,below=1pt,fill=white]{direction of light propagation} ;
\end{scope}
\end{tikzpicture}
\end{figure}

The physical setup of a non-linear curvature wavefront sensor (nlCWFS) is a little complicated however, for the purposes of this phase-retrieval algorithm, it can be considered to consist of measurement planes located at distances $\pm z_1$ and $\pm z_2$ from the pupil plane, as shown in \fref{fig:planediagram}.

Unless otherwise indicated, all calculations are based on a $D=4.2\m$ annular aperture with inner diameter $1\m$ and $z_1$ and $z_2$ are $1000\km$ and $2500\km$ respectively.  All image data are assumed to be  256 pixels square. These values correspond to the William Herschel 4.2m telescope on La Palma in the Canary Islands.  For Lucky Imaging (LI)  to be a useful technique, the probability $p_{Lucky}$ of a lucky exposure (defined here as having RMS phase error across aperture less than 1 radian) needs to be high enough that it might be achieved many times within one observing session. Fried's expression for $p_{Lucky}$ is:
\begin{equation}
p_{Lucky} \simeq 5.6 \exp\left[-0.1557\left(\frac{D}{r_0}\right)^2\right]\;
\end{equation}
where $r_0$ is the Fried parameter. 

Lucky Imaging has been shown to work well on $2.5\m$ telescopes, for which, assuming $r_0=0.35\m$ yields $p_{Lucky}=0.002$.
The target, then, for the algorithms below is to restore $p_{Lucky}$ to this range for larger telescopes. This can equivalently be stated as reducing the effective number of turbulent cells (across which the RMS phase error is 1 radian)  across a diameter to approximately 8, or, alternatively, increasing the effective value of $r_0$ to $\frac{D}{8}$.

Whichever analogy is used, the requirement is that the RMS phase error is reduced to around $1\rad$ in some non-negligible proportion of exposures.

To represent phase screens mathematically, the screen is broken up into individual pixels each represented by a complex number. Light with phase angle $\phi$ and amplitude $A$ is represented by the complex number $A\,e^{i\,\phi}$. To recover the optical path difference (OPD), which is needed to drive deformable mirrors for adaptive optic correction, the phase must be unwrapped, by adding and subtracting multiples of $2\pi$ where appropriate to give a smooth function without discontinuities. The pixel spacing in these simulations is equivalent to $d=0.05\m$ in the telescope pupil and the wavelength $\lambda = 750\nm$. Input phase is generated from an implementation by \citet[170]{NumSim} of a technique described by \citet{Harding99} for generating phase screens from the Kolmogorov model of turbulence, \autocite{KolmogA,KolmogB} using $L_0=100\m$, $l_0=0.01\m$ and $r_0=0.5\m$.   These correspond to the outer scale size, the inner scale size and the turbulent uncorrected cell size. 

The value of $r_0$ represents an unrealistic scenario for uncorrected wavefronts. It was not chosen through optimism but for computational simplicity as the rudimentary phase unwrapping process often failed (even for well-reconstructed wavefronts) with lower values of $r_0$, rendering meaningful algorithm comparison difficult. The mean RMS optical path difference in the input was approximately 500nm, which represents a realistic scenario of what might happen when running the AO system in `closed loop' configuration. The algorithm did successfully converge with larger aberrations (lower $r_0$) and theoretical calculations suggest that the algorithm ought to provide useful output for values of $r_0$ around $10\cm$ (see \fref{fig:gridconstraint}).

The  proposed optical setup  involves splitting the beam  into 4 different colour bands using dichroics (so each near-pupil plane image in fact is formed through a different pass band), but this  is not important for the consideration of this paper. It is of note, however, that as a result of the efforts made in instrumentation to ensure the achromaticity of the optical system, the light is assumed to be of one wavelength only. Our simulations suggest that this is a good approximation particularly for the low order corrections that we wish to achieve.


\section{Algorithm}
\label{sec:alg}

\subsection{Summary of algorithms}
\label{sec:alg:sum}

The iterative algorithms presented here share the same broad structure. The inputs are $m_p$, the measured amplitudes at each plane $p_i$, and, optionally, $g_{0}(p_i)$, the initial phase candidates.

Conceptually, the algorithms involve using known constraints to recursively refine an estimate of the pupil phase. These constraints are the measured intensities in the four measurement planes and the pupil shape.  An accurate estimate for the pupil phase must reproduce (upon simulated propagation) the measured intensities at the four planes, with zero intensity outside the aperture of the pupil. It may be possible to assume uniform pupil illumination and use this as a further constraint, although this is likely to present problems at low photon counts, when the pupil will have few enough points illuminated that the assumption of uniformity breaks down. 

The algorithm begins by making an estimate of the current pupil phase. A  numerically simulated propagation is then performed, to give an estimate for the phase and amplitude of the wavefront at one of the measurement planes based on the pupil estimate. The amplitude estimate at each pixel is then discarded and replaced with the amplitude derived (by square-rooting) from the measured intensity. The phase estimate is retained. This new estimate is propagated to another measurement plane, where again the phase estimate is retained, whilst replacing the estimated amplitude with the measured one. This process of propagating between planes is repeated until a given number of iterations have been performed, or until the estimate has completely succeeded (or completely failed) to converge.

Mathematically, the algorithms are described as follows, where $A_{i,j}$ indicates a propagation from plane $i$ to plane $j$:
\begin{align}
  g^{\prime}_{k}(p_j) &=A_{i,j} \left[g_{k}(p_i) \right]\\
  g^{\prime\prime}_{k}(p_j) &=\sum\limits_{m=k}^0 a_{j,(k-m)}\,g^{\prime}_{m}(p_j) + b_{j,(k-m)}\,g_{m}(p_j)\\
  g_{k+1}(p_j) &= \arg\left[g^{\prime\prime}_{k}(p_j)\right]\,\abs\left[m_{p}\right]
\end{align}
Most of the $a_i$ and $b_i$ are zero, and, since multiplying each $a_i,\,b_i$ by a constant has no effect, we place an arbitrary restraint that
\begin{equation}
\sum(a_i + b_i)=1\;.
\label{restraint}
\end{equation}
In \tref{tab:algsum} and all subsequent descriptions, only the non-zero $a_{i}, b_{i}$ are given.

\begin{table}[h]
\caption{Summary of parameter values for different algorithms. $h$ is a feedback parameter and represents the fraction of the phase used from input in the next estimate.}
\label{tab:algsum}
\begin{center}
\begin{tabular}{ l l }
\toprule
Name & Parameters\\
\midrule
Error-Reduction & $a_0=1$ \\
   Input-Output & $a_0=h, b_0=1-h$ \\
  Output-Output & $a_0=1+h, a_1=-h$ \\
     In-Out-Out & $a_0=h, a_1=-h, b_0=1$ \\
\bottomrule
\end{tabular}
\end{center}
\end{table}

The approach of Fienup (\citeyear{Fienup82}) was only designed for use in the `linear' region of the propagated wavefront, defined as the region where a small change in phase at a certain point in the `input' gives rise to a small change in the phase at the corresponding point in the `output' and nowhere else. The algorithm which has been developed here works in the non-linear region, despite not being designed to do so, although a formal analysis of why this is the case is not the subject of this paper.

\subsection{Algorithms for use in four planes}

In extending the standard algorithms to four planes, several variations are possible, such as the order in which propagation is done between the planes. Clearly, with two planes, there is only one scheme possible, but, eliminating cyclic and reflectional symmetries in the visiting order, with four planes there are $4!/(4\cdot2)=3$ possible visiting cycles, as shown in \tref{visits}. 
\begin{table}[h]
\caption{Visiting orders for four planes}
\label{visits}
\begin{center}
\begin{tabular}{@{Scheme } c @{: ( }c @{ , } c @{ , } c @{ , } c @{ ) }}
1 & 1 & 2 & 3 & 4\\
2 & 1 & 2 & 4 & 3\\
3 & 1 & 3 & 2 & 4\\
\end{tabular}
\end{center}
\end{table}

Since pairs $( p_1, p_2)$ and $( p_3, p_4)$ shown in Figure 1 contain similar length-scale information, it  can be assumed that Schemes 1 and 2 will be broadly similar. However, we might expect Scheme 3 to behave slightly differently, as each propagation moves to a plane with different length-scale information.

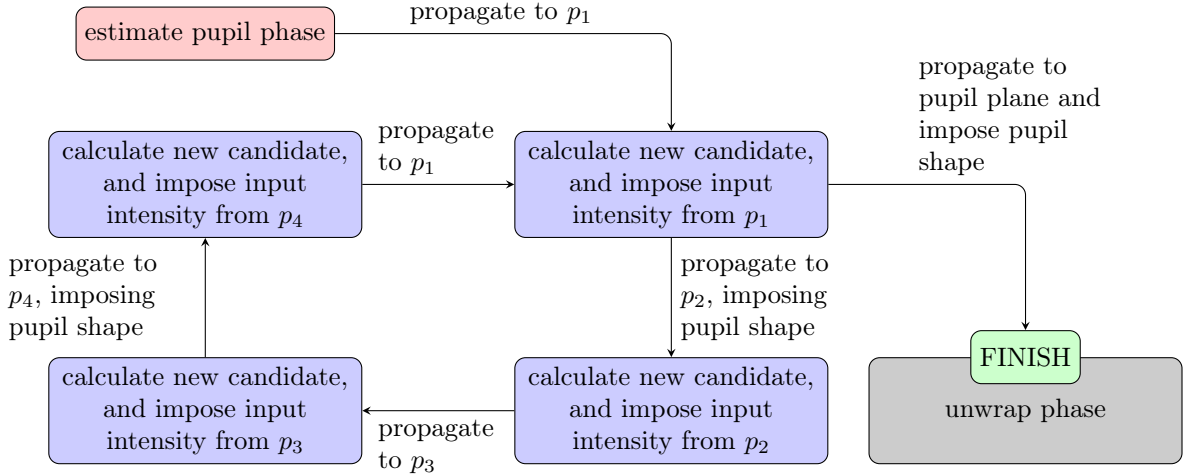
\begin{figure*}
\caption{Schematic diagram of the four-plane reconstruction method. The step in which the new candidate is calculated is different in each algorithm.}
\label{schematic}
\begin{tikzpicture}[node distance = 3.0cm, auto]
    \node [plane] (p1) {calculate new candidate, and impose input intensity from $p_1$};
    \node [stage] at ([shift=({0.6\columnwidth,-3cm})]p1) (unwrap) {unwrap phase};
    \node [finish] at (unwrap.north)  (finish) {FINISH};
    \node [plane, below of=p1] (p2) {calculate new candidate, and impose input intensity from $p_2$};
    \node [plane, left=2cm of p2] (p3) {calculate new candidate, and impose input intensity from $p_3$};
    \node [plane, above of=p3] (p4) {calculate new candidate, and impose input intensity from $p_4$};
    \node [start, above of=p4, node distance=2cm] (start) {estimate pupil phase};
    \path [line] (start)  -| node[plabel, near start ]{propagate to $p_1$}  (p1);
    \path [line] (p1)  edge node[plabel]{propagate to $p_2$, imposing pupil shape} (p2);
    \path [line] (p2)  edge node[plabel2]{propagate to $p_3$} (p3);
    \path [line] (p3)  edge node[plabel]{propagate to $p_4$, imposing pupil shape} (p4);
    \path [line] (p4)  edge node[plabel2]{propagate to $p_1$} (p1);
    \path [line] (p1)  -| node[plabel1, above]{propagate to pupil plane and impose pupil shape}  (finish);
\end{tikzpicture}
\end{figure*}

A second possible variation is the choice of what type of `step' is applied at each plane. The Input-Output (IO) algorithm starts with an estimate of the input phase.    This is then propagated to generate a phase at the output position.  This output phase will then be used for the next iteration generally combined with some fraction of the input phase.  It is possible to apply an `input-output' step after four propagations (i.e. once every iteration), after every two propagations (twice per iteration), or after every iteration (four times per iteration). These different algorithms will be referred to as IO1, IO2, and IO4 respectively, and use a similar naming convention for the Output-Output (OO) and In-Out-Out (IOO) algorithms. In each case, all `ordinary' steps are ER steps. Furthermore, it is  possible to use a different value of $h$ at the different planes, for example $h=h_1$ at $p_1$ and $p_2$, $h=h_2$ at $p_3$ and $p_4$.


\subsection{Effect of arbitrary phase difference}
\label{sec:alg:arbphase}

The algorithms will, of course, recover only the relative phase, as there is no reference phase information available. Therefore, an arbitrary phase difference --- generally different for each iteration, but constant across the plane --- will be present in each algorithm at each plane.

In the development of these algorithms, only a basic attempt has been made to correct for the arbitrary phase difference introduced at each plane. Each of the initial estimates at the measurement planes has been assigned a uniform phase angle by multiplying the zero-phase values by $e^{i k z}$, where $k = 2\pi /\lambda$ is the wavenumber. With an ER algorithm, the overall phase difference between iterations is unimportant --- the phase of the previous iteration is discarded and replaced. When the difference between iterations is used to calculate the new phase, however, (i.e. all algorithms except ER) a problem arises if there is an additional arbitrary phase $\phi_a$.

The direction of the propagation chosen will be influenced by this arbitrary phase: whether this is catastrophic for the convergence of the algorithm depends on the probability distribution of $\phi_a$. If $\phi_a \sim \rand{-\pi}{\pi}$ (the phase angle introduced between iterations is distributed uniformly over all angles), the reconstruction will become a random walk for some (larger) values of $h$; adding a uniformly distributed random angle to any angle yields a uniformly distributed random angle, due to the fact that the angle `wraps around'. The phase error at each step due to this arbitrary phase can be found by comparing the phase angle with and without it. The calculation of this error $\phi_e$ for the IO algorithm is as follows. The assumption below is that $A_{k}\simeq A_{k+1}$, where $A_{k}$ is the amplitude of a given pixel at the $k^{th}$ iteration. $\phi_c=\phi_{k+1}-\phi_k$ is the phase correction at the $k^{th}$ iteration.

\begin{figure*}[tbp]
\caption{Progress of IO4 ($h=1.1$) for 30 iterations. Determination of $\phi_a$ and $\phi_c$ was at $p_1$.}
\label{fig:phi_a}
\begin{tikzpicture}
 \begin{semilogyaxis}[
 	width=1.8\columnwidth,
	height=0.8\columnwidth,
	xlabel={Iteration},
	xmin=0,
	xmax=30,
	ylabel={Phase angle $/ \rad$}
]
\pgfplotstableread{phi_a.dat}\phia

\pgfplotsset{cycle list name=linestyles}

\addplot table[x index=0,y index=1] \phia; \addlegendentry{$\abs(\phi_a)$};
\addplot table[x index=0,y index=2] \phia; \addlegendentry{RMS $\phi_c$};
\addplot table[x index=0,y index=3] \phia; \addlegendentry{RMS residual};
 \end{semilogyaxis}
\end{tikzpicture}
\end{figure*}
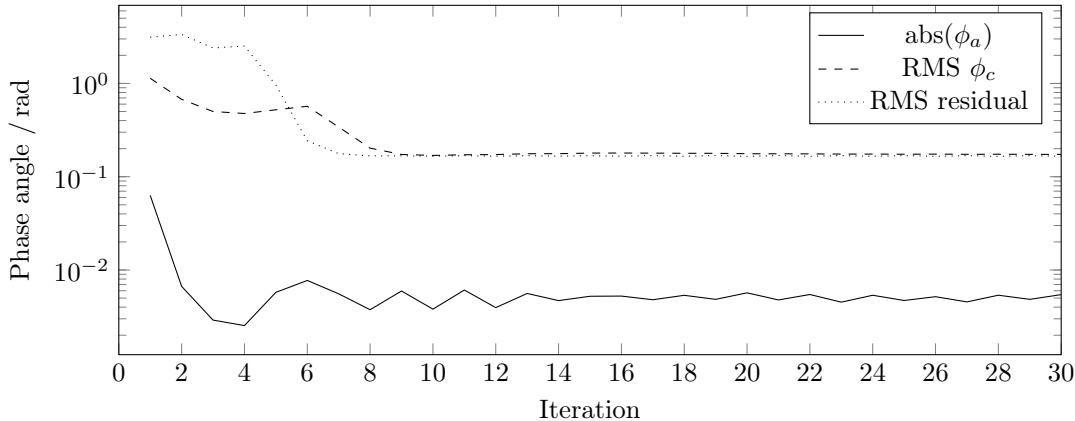

For the effect of the arbitrary phase difference not to hamper the convergence of the algorithm, the phase must be distributed more tightly than \rand{-\pi}{\pi}. Inspection of the function in \eqref{error_z} reveals that for good convergence, either $b$ or $\phi_a$ must be small, the latter in the `small angle approximation' sense. It would appear that in most cases, the requirement holds, or at least approximately holds, but it is possible to directly test the distribution of $\phi_a$, and to experimentally compare its magnitude to $\phi_c$ --- see \fref{fig:phi_a}. An estimation of $\phi_a$ is possible, if it is small, by taking the mean of the phase difference between iterations, weighted by the intensity of the signal: 
\begin{equation}
\phi_a\simeq\frac{\sum_{j}(\phi_{(k+1),j}-\phi_{k,j})I_j}{\sum_{j}I_j}
\end{equation}

\begin{align}
\phi_e &=\arg[z]\textrm{, where}
\\
z&=\frac{e^{i \phi_a}\left(e^{i \phi_k}+ h \left(e^{i \phi_k + i \phi_c}- e^{i \phi_k}\right)\right)}{e^{i \phi_k}+ h \left(e^{i \phi_k + i\phi_c +i\phi_a}- e^{i \phi_k}\right)}
\nonumber\\
&=\frac{e^{i \phi_a}\left(1+ h \left(e^{i \phi_c}- 1\right)\right)}{1+ h \left(e^{i \phi_c + i\phi_a}- 1\right)}
\quad[\textrm{let $b=h-1$}]
\nonumber\\
&=1 +\frac{b \left(1-e^{i \phi _a}\right)}{e^{i \phi _a+i \phi _c}+b \left(e^{i \phi _a+i \phi _c}-1\right)}
\nonumber\\
&=1+e^{-i(\phi_a+ \phi_c)}\frac{b \left(1-e^{i \phi _a}\right)}{1 + b \left(1-e^{-i(\phi_a+\phi_c)}\right)}
\nonumber\\
&=1+e^{-i(\frac{\phi_a}{2}+ \phi_c)}\frac{ -2\,i\,b\sin\left(\frac{\phi_a}{2}\right)}{1 + b \left(1-e^{-i(\phi_a+\phi_c)}\right)} \label{error_z}
\end{align} 

Correcting for this phase, therefore, may help in improving the stability of the algorithm at higher values of $h$ --- in practice, however, calculation of $\phi_a$ must be computationally efficient and quick, or fewer iterations of the algorithm will be possible in a preordained time, potentially negating the benefits of its correction. It is also possible that, given the simple scheme of calculating $\phi_a$, assuming $\phi_{(k+1),j}-\phi_{k,j}=\arg\left[e^{i(\phi_{(k+1),j}-\phi_{k,j})}\right]$, (i.e. without phase unwrapping), a misestimation is likely, which could increase the error it seeks to avoid. Consider the scenario of an arbitrary phase of $\phi_a \sim\pi$ added to actual corrections of order $\sim0.1$. The calculated value of $\phi_a$ will be $\sim0$ if the real value of $\phi_a$ is sufficiently close to $\pi$. Fortunately, the actual value of $\phi_a$, as can be seen from \fref{fig:phi_a}, is an order of magnitude lower than $\phi_c$ in the case with ample light. Assuming this size difference is similar at low light-levels, correcting for $\phi_a$ will not be of great importance for algorithm performance.

\subsection{Propagation Techniques}
\label{sec:alg:proptech}

The Fresnel integral describing the propagation of a known wavefront through space can be expressed in the form of a two-dimensional Fourier transform and two multiplications. Discretising the Fourier transform and using Fast Fourier Transform (FFT) techniques gives a computationally efficient way of simulating the propagation of a wavefront between two parallel planes. The discrete nature of the calculation fixes the relative grid spacing of the sampling at the two planes so in order to use this simple one-step technique, (so-called because it only requires one FFT) the grid spacings of the planes must have a particular relationship, or the distance between them must have a specific value. The grid spacings of the measurement planes is fixed by the pixel size of the imaging equipment, so the distance between planes is fixed. There is no guarantee that arranging the measurement planes at these precise relative distances gives good performance in a reconstruction algorithm.

In \citeauthor{NumSim}'s book, two more flexible numerical propagation techniques are outlined \autocite[Chapter 6]{NumSim}. \citeauthor{NumSim} explains four theoretical sampling constraints, determining the limiting relationships between grid size, grid spacings at each plane, and propagation distance.  

The first technique the two-step Fresnel propagation, involves splitting the required propagation into two steps, each step involving one Fourier transform (FT). When, as here, the grid spacings of the two planes are equal, the technique is equivalent to propagating from the first plane to a plane half way between the first and second planes, and then on to the second plane, with the grid-spacing at the intermediate plane being chosen to satisfy the one-step constraint on distance and grid-spacing. Because the CWFS planes are symmetric about the pupil plane, when propagating from $p_1$ to $p_2$, or from $p_3$ to $p_4$ (see \fref{schematic}), this intermediate plane is the pupil plane, albeit with a different grid-spacing from the other planes. Using  schemes 1 or 2 (see \tref{visits}),  allows us to impose the pupil shape constraint twice per iteration with no extra FTs. To do this would otherwise add 50\% to the running time of each iteration.

The second technique, the angular spectrum method, also involves two FTs, but here there is no intermediate plane, so it is impossible to implement any clever tricks and impose the pupil shape `for free'. Additionally, the sampling constraints to avoid frequency aliasing are such that the technique works best only for small propagation distances, whilst the Fresnel method works best only for large distances. Because the power spectrum of the phase error due to atmospheric turbulence has a greater value at lower frequencies. The high-order structures which appear in the working region of the angular spectrum method will contribute much less to the error than the low-order structures visible in the working region of the two-step Fresnel method. The exception to this rule is when extremely coarse binning is used, such as may be the case at low light-levels (see,  \sref{sec:lowlight}).

Because of these effects, algorithms use Fresnel propagation and visiting scheme 1 in \tref{visits}, as the speedup obtained from imposing the pupil shape `for free' is more effective than any possible faster convergence from visiting the planes in a different order. The implementation presented here requires eight $256\times256$ FTs for each iteration (two per propagation), and it is this which is expected to be responsible for the majority of the execution time. If, as is seen in some instances, an algorithm converges to a solution in 5 iterations, this requires $40\;$ $256\times256$ FTs.

\subsection{Grid spacing}
\label{sec:alg:grid}

Unlike with a SHWFS, when using  a nlCWFS each grid point in the pupil plane holds phase information, rather than phase slope information for each sub-aperture. As a result of this, once the true (i.e. unwrapped) phase difference between a region of neighbouring grid points is greater than $\pi$, any phase-unwrapping algorithm, or even any phase-slope fitting method, will struggle to identify the true unwrapped phase.


The expected value of the squared phase difference between two points is expressed \autocite{Fried65} in terms of the Fried parameter $r_0$ as:
\begin{equation}
D_\phi\left(\left|\vect{r}_1-\vect{r}_2\right|\right)=6.88\left(\frac{\left|\vect{r}_1-\vect{r}_2\right|}{r_0}\right)^{5/3}
\end{equation}

This expression represents an expectation value --- the maximum value may be much larger, but is less likely to occur. If the errors in the reconstruction algorithm add a phase error at each point that is normally distributed about 0 with variance $\sigma^2=\theta_e^2$, and the true phase difference between grid points is normally distributed about 0 with variance $\sigma^2=D(d)$, then the reconstructed phase difference between adjacent grid points will be distributed with mean $0$ and variance $\sigma^2=2\theta_e^2+D(d)$. The error in the algorithm will cause wrap-around between pixels with probability $p_{wrap}$ if:
\begin{equation}
2\left(\mathrm{erfc}^{-1}(p_{wrap})\right)^2\left(2\theta_e^2+D(d)\right)>\pi^2\;.
\end{equation}
where erfc is the complementary error function. The grid-spacing $d$ permissible is then given by:
\begin{eqnarray}
6.88\left(\frac{d}{r_0}\right)^{5/3}+2\theta_e^2<\frac{1}{2}\left(\frac{\pi}{\mathrm{erfc}^{-1}(p_{wrap})}\right)^2 \nonumber \\
d < 0.314\, r_0 \,\left[\frac{1}{2}\left(\frac{\pi}{\mathrm{erfc}^{-1}(p_{wrap})}\right)^2-2\theta_e^2)\right]^{3/5}
\end{eqnarray}

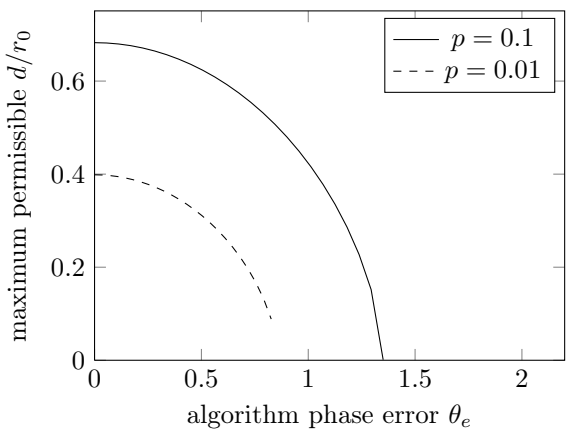
\begin{figure}[h]
\caption{Constraint on $d$ needed to avoid wrap-around in the recovered phase}
\label{fig:gridconstraint}
\begin{tikzpicture}
\begin{axis}[
 width=\columnwidth,
 height=0.8\columnwidth,
 xlabel={algorithm phase error $\theta_e$},
 ylabel={maximum permissible $d/r_0$},
 xmin=0,
 xmax=2.2,
 ymin=0
]
\pgfplotsset{cycle list = {black}}
\addplot +[domain=0:1.35064] {0.314*(((1/2)*(pi/1.163)^2-2*x^2)^(3/5))}; \addlegendentry{$p=0.1$}
\pgfplotsset{cycle list = {dashed}}
\addplot +[domain=0:0.8627] {0.314*(((1/2)*(pi/1.821)^2-2*x^2)^(3/5))}; \addlegendentry{$p=0.01$}
\end{axis}
\end{tikzpicture}
\end{figure}

This constraint on $d$ is shown in \fref{fig:gridconstraint}. In a closed loop situation, this condition is less restrictive, as the effective $r_0$ is increased.

The error-dependence of this constraint is unsurprising --- we would reasonably expect that any error in the reconstruction must be compensated by a finer grid of sampling points --- but it does provide a useful upper bound on $d$ in terms of $r_0$. For example, with an effective closed-loop Fried parameter of $1\m$, an error of  around $0.7\rad$ in the reconstruction algorithm would allow a grid spacing of 20cm if 1\% was an acceptable likelihood of wrap-around effects.

Clearly the spatial extent of the grid must be at least as large as the size of the pupil, or not all phase information will be recovered. An aberrated wavefront will spread spatially on propagation. A finite fourier transform makes the assumption that the image is periodic on its boundaries. For the finite size of the grid not to cause overlapping (``wraparound'') at the edges of the propagated images, the pupil plane image must therefore be padded with zeros. We define the padding factor to be $M_{padding}=\frac{Nd}{D}$ --- this is the linear ratio of the grid extent to the pupil extent --- and choose a value such that the linear grid size $N$ is a power of two. 

\subsection{Parallelisation}
\label{sec:alg:para}

Two-dimensional FFTs can be effectively parallelised and executed on  Graphics Processor Units (GPUs). The use of GPUs opens the additional possibility of using a parallelised phase retrieval algorithm. In this case, the flow of execution is somewhat more complicated, and many possibilities are available. A basic approach has been explored here.

\begin{figure*}
 \caption{Conceptual diagram of a parallelised algorithm. Arrows represent propagations. The vertical direction represents both time (increasing downwards) and error (decreasing downwards, theoretically). Each column represents a plane, and each of the four threads execute only one type of propagation (e.g. propagating from $p_1$ to $p_2$) and the subsequent $h$-sized step, indicated by a dotted line.}
 \label{fig:para}
 \centering
\begin{tikzpicture}[>=stealth]


\node (pl1) [pl,fill=red!20] {};
\node (pl2) [pl,fill=green!20,anchor=west]  at (pl1.east) {};
\node (pl3) [pl,anchor=west]  at (pl2.east) {};
\node (pl4) [pl,fill=orange!20,anchor=west]  at (pl3.east) {};


\node (l1) [pllabel] at (pl1.north) {$p_1$};
\node (l2) [pllabel] at (pl2.north) {$p_2$};
\node (l3) [pllabel] at (pl3.north) {$p_3$};
\node (l4) [pllabel] at (pl4.north) {$p_4$};

\node (i11) [inp,below=1em of l1] {};
\node (i12) [inp,below=1em of l2] {};
\node (i13) [inp,below=1em of l3] {};
\node (i14) [inp,below=1em of l4] {};
\node (o11) [outp,below=1em of i11] {};
\node (o12) [outp,below=1em of i12] {};
\node (o13) [outp,below=1em of i13] {};
\node (o14) [outp,below=1em of i14] {};
\node (i21) [inp,below=2em of o11] {};
\node (i22) [inp,below=2em of o12] {};
\node (i23) [inp,below=2em of o13] {};
\node (i24) [inp,below=2em of o14] {};
\node (o21) [outp,below=1em of i21] {};
\node (o22) [outp,below=1em of i22] {};
\node (o23) [outp,below=1em of i23] {};
\node (o24) [outp,below=1em of i24] {};
\node (i31) [inp,below=2em of o21] {};
\node (i32) [inp,below=2em of o22] {};
\node (i33) [inp,below=2em of o23] {};
\node (i34) [inp,below=2em of o24] {};
\node (o31) [outp,below=1em of i31] {};
\node (o32) [outp,below=1em of i32] {};
\node (o33) [outp,below=1em of i33] {};
\node (o34) [outp,below=1em of i34] {};

\draw [->] (i11)--(o12);
\draw [->] (i12)--(o13);
\draw [->] (i13)--(o14);
\draw (i14)-- +(0.25\columnwidth,-0.5em-0.75ex);
\draw [<-] (o11) -- +(-0.25\columnwidth,0.5em+0.75ex);

\draw [->] (i21)--(o22);
\draw [->] (i22)--(o23);
\draw [->] (i23)--(o24);
\draw (i24)-- +(0.25\columnwidth,-0.5em-0.75ex);
\draw [<-] (o21) -- +(-0.25\columnwidth,0.5em+0.75ex);

\draw [->] (i31)--(o32);
\draw [->] (i32)--(o33);
\draw [->] (i33)--(o34);
\draw (i34)-- +(0.25\columnwidth,-0.5em-0.75ex);
\draw [<-] (o31) -- +(-0.25\columnwidth,0.5em+0.75ex);


\draw[dotted]
(o11)--(i21)
(o12)--(i22)
(o13)--(i23)
(o14)--(i24)

(o21)--(i31)
(o22)--(i32)
(o23)--(i33)
(o24)--(i34);

\end{tikzpicture}
\end{figure*}
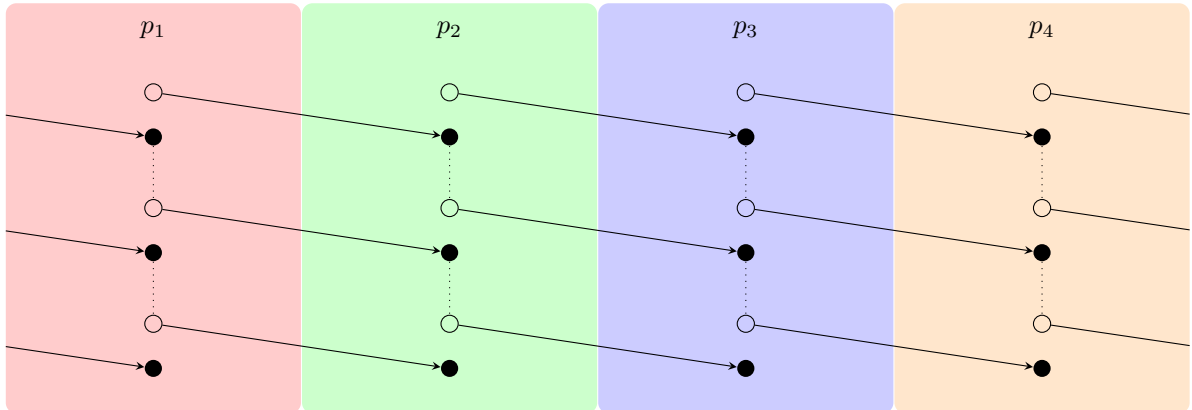

Because of the serial nature of text, this concept is best explained using a diagram (\fref{fig:para}).


When the parallel algorithm terminates, a choice of four different outputs (each of the four candidate solutions, one from each measurement plane) is available. Using all four estimates of the pupil phase and combining information from them for phase recovery is likely to result in a better estimate being achieved.

\section{Low light levels}
\label{sec:lowlight}

At high frame rates and with faint reference stars, low photon counts in the imaging planes may be expected. Typically, data binning is used when light levels drop, however, the sampling constraints in the Fourier domain for the Fresnel propagations used in the algorithms mean that to reduce the grid-size used in propagations, the measurement planes must be at greater distances from the pupil plane. The distance $\Delta z$ of the measurement plane from the pupil plane must satisfy:
\begin{equation}
\Delta z \geq \frac{(D \, M_{padding})^{2}}{\lambda \, N}\quad ,
\label{distfrompupil}
\end{equation}
where $D$ is the aperture diameter, $M_{padding}$ is the padding factor, $\lambda$ is the wavelength, and $N$ is the grid size \autocite{NumSim}. The angular spectrum propagation technique  must be used in this case. Even if the angular spectrum method is used, spatial binning still presents a fundamental problem for accurate reconstruction of the OPD (see \sref{sec:alg:grid}).

\subsection{Gaussian convolution}
\label{sec:lowlight:GC}

Signals at low light levels are dominated by Poisson noise. Because of this,  rather than binning the data we preprocess it by convolving it with a 2D Gaussian function (essentially `blurring' the image). Gaussian convolution (GC) has the advantage of preserving the spatial information that would be lost by binning, and ensures that all pixels have non-zero intensities. Binning can still leave some pixels with zero intensity.

The speckle size at distance $z$ from the pupil plane is given by $\frac{\lambda z}{D}$, so a GC of around this size could reasonably be expected to perform well. Since $z$ is different at each plane, it would make sense to use a different sized GC for planes at $\pm z_2$ from $\pm z_1$, which will  be referred to as DGC (differential GC). Both of these techniques require the execution of two FFTs.

\subsection{Triangular Filling}
\label{sec:lowlight:awtf}
With a grid size of $256\times256$ pixels, and low photon counts of $O(10^4)$ or fewer (i.e. fewer photons than pixels), we can expect that the majority of pixels will receive no photons, and a small number will receive one photon (possibly more). We need to infer intensity from photon density in this region. Techniques exist for this purpose \autocite{WilletNowak04}, although many assume regions of uniform intensity, separated by smooth boundaries, which is not true of the intensity measurements we expect to make at low light levels. We propose the following approach be used at each measurement plane at extremely low photon counts:
\begin{enumerate}
\item Let $\cal P$ be the set of pixels with non-zero intensities.
\item Compute $\cal T$, the Delaunay triangulation of $\cal P$.
\item For each triangle in $\cal T$, set pixels of this triangle to $\frac{1}{A}\sum\limits_{n=1}^{3} I_{n}w_{n}$, where $I_n$ is the value at vertex $n$, $w_n$ is a weighting factor, equal to the portion of the pixel which the triangle covers, and $A$ is the pixel area of the triangle.
\end{enumerate}
This technique of area-weighted triangulation filling (AWTF) should be well-suited to GPU implementation. Rendering triangle meshes is a task that GPUs are well-designed for, although the time taken depends upon options such as anti-aliasing techniques as this has not been verified experimentally. Delaunay triangulation for $n$ points can be performed in $O(n \log n)$ time.  Fortunately the scenario in which AWTF is needed the most (fewest non-zero pixels) is also the scenario in which it executes fastest. AWTF can be used in conjunction with GC, in that order.

\subsection{Voronoi Filling}
\label{sec:lowlight:awvf}
Another polygon-based technique may have even better performance, however: the Voronoi diagram or tesselation of the non-zero points. Voronoi tesselation has been used in image reconstruction in conjunction with Bayesian methods \autocite{Cabrera08}. The method presented here is somewhat simpler than that presented by \citeauthor{Cabrera08} however, in that the non-zero pixels are used to compute the tesselation, rather than any more complicated method. The steps used to compute the area-weighted Voronoi filling (AWVF) are:
\begin{enumerate}
\item Let $\cal P$ be the set of pixels with non-zero intensities.
\item Compute $\cal V$, the Voronoi tesselation of $\cal P$ (bounded by the image limits).
\item For each polygon $p$ in $\cal V$, set pixels within this polygon to $\frac{I}{A}$, where $I$ is the intensity of the pixel inside $p$ and $A$ is the pixel area of $p$.
\end{enumerate}
Voronoi filling has the unfortunate property of removing all of the localisation (of intensity) from the image, and this causes it to be a poor technique to use on its own. Using functional notation, the following scheme somewhat mitigates this problem:
\begin{equation}
\sqrt{\mathrm{DGC4}(\mathrm{AWVF}(I))} \cdot \mathrm{DGC4}(I)
\label{eq:vor}
\end{equation}
The approach shown in \eqref{eq:vor} , which will be referred to as Gaussian-Voronoi (GV), has the advantage of preserving the intensity localisation, but adjusts the intensity proportionally to the linear density of observed photons (the inverse of the distance between them).

\subsection{Guiding Lucky Imaging selection in low light}
\label{sec:lowlight:LIselection}

Lucky Imaging requires the selection of images based on some criteria of sharpness. If the reconstruction of the pupil phase is poor, the AO correction applied will not adequately correct the atmospheric error and the science image will not be sharp. Measures available during the convergence of the algorithm can give an indication of how successful convergence has been. These measures may then be used to guide the selection of images for combining using LI techniques. One such measure is the intensity-weighted variance of the phase difference between successive iterations of the algorithm  at any of the measurement planes. If this is high, the convergence is either poor or unfinished. This measure could either be used during capture or retrospectively to select exposures for LI.

\subsection{Comparison of low-photon-count strategies}
\label{sec:lowlight:compstrat}


\begin{figure*}[htb]
  \caption[Pictures of PSFs]{Central portion of \subref{fig:psf_ideal} the ideal Point Spread Function, and the PSF of  \subref{fig:psf_raw} the uncorrected wavefront, and wavefronts corrected using \subref{fig:psf_dgc8} DGC8 + 5, \subref{fig:psf_awtf} AWTF, DGC4  + 5, \subref{fig:psf_gv} GV + 4, where `+ $n$' denotes $n$ iterations of IO4 with $h=1.08$. RMSE in the input was $590\nm$ and there were 393 photons.}
  \label{fig:psfs}
  \centering
  \subfloat[]{\label{fig:psf_raw}\includegraphics[width=\picwidth]{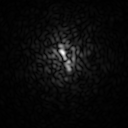}}
  \hspace*{0.1em}
  \subfloat[]{\label{fig:psf_dgc8}\includegraphics[width=\picwidth]{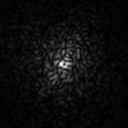}}
  \hspace*{0.1em}
  \subfloat[]{\label{fig:psf_awtf}\includegraphics[width=\picwidth]{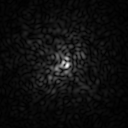}}
  \hspace*{0.1em}
  \subfloat[]{\label{fig:psf_gv}\includegraphics[width=\picwidth]{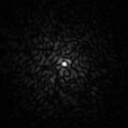}}
  \hspace*{0.1em}
  \subfloat[]{\label{fig:psf_ideal}\includegraphics[width=\picwidth]{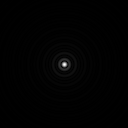}}
\end{figure*}


In simulations of various light levels, the same reconstruction strategy --- 30 iterations of the IO4 algorithm with parameter $h=1.08$ --- was used for all inputs. 100 independent phase screens as described in \sref{sec:instr} were used to generate complex pupil functions, which were then propagated to the four image planes. Poisson noise was then added to the intensity (amplitude squared) images, which was used as the input to the correction techniques or to the algorithm directly. The resulting pupil phase was then unwrapped and the RMS error (i.e. the RMS value of the difference between the unwrapped reconstructed phase and the input phase) was computed. Naturally this incorporates into the results the successes (and failings) of the unwrapping algorithm, although comparing the wrapped phases would limit the range of error values to $\pm\pi\rad$. It is therefore preferred to the alternative of no unwrapping. In the low photon-count limit however, the unwrapping stage frequently partially failed, due to lack of sharpness of the phase in the pupil plane. This in turn gave a high RMS error due to the spurious jumps of multiples of $2\pi$. In these instances, the output RMS error was higher than the RMS error of the input, which is clearly not desirable or representative. Using a more robust unwrapping technique or a Zernike phase-slope technique might allow unwrapping to be successful at low photon counts, enabling further study of this region. By dividing the aberrated complex pupil function by the reconstructed complex pupil function, clamping the amplitude to 1 for all pupil plane pixels, and calculating the resulting point spread function (PSF), a somewhat clearer indication of the true performance is available compared with  what can be obtained by the limited unwrapping techniques available. A comparison of the central portions of PSFs for a randomly generated input phase is shown in \fref{fig:psfs}. Here, the relative performance of the low light techniques is clearly visible.

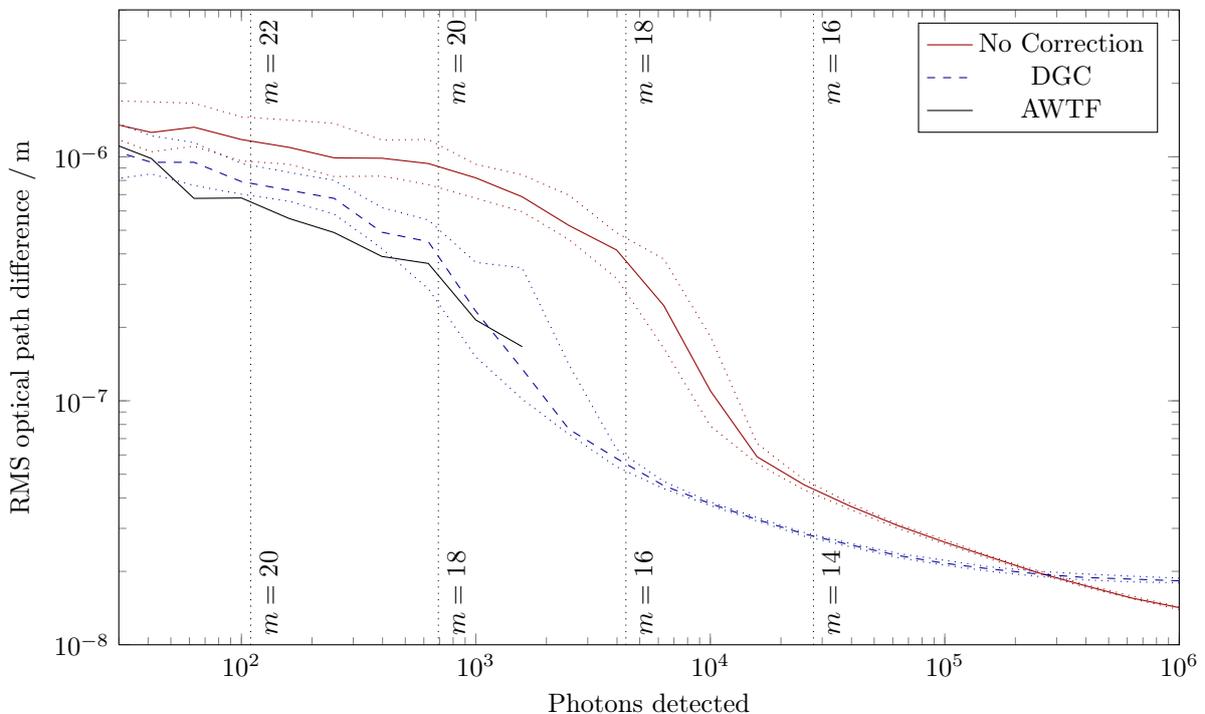
\begin{figure*}[htpb]
\caption{Simulated low light performance of CWFS using algorithm IO4 with and without preprocessing the input planes with DGC3 and AWTF (see text). Dotted lines show the upper and lower quartiles. AWTF is only used below 2000 photons. Vertical lines show the I-band magnitude required for a given photon count with the CWFS running at $10\hz$ for $D=4.2\m$ (lower label), and $D=10.5\m$ (upper label).}
\label{fig:strats}
\begin{tikzpicture}[node distance = 0.05cm]
\begin{loglogaxis}[
    width=2\columnwidth,
    height=10cm,
    ylabel={RMS optical path difference $/\m$},
    xlabel={Photons detected},
    xmin=30, xmax=1000000,
    ymin=1e-8,ymax=4e-6,
    minmax/.style={dotted,mark=none,forget plot},
    quartile/.style={dotted,mark=none,forget plot},
    med/.style={solid,mark=none}
]

\pgfplotsset{cycle list={{red!60!black}}}

\pgfplotstableread{100t0b0.dat}\raw
\pgfplotstableread{100t0b3i1_5.dat}\blura
\pgfplotstableread{100t1b0.dat}\rawt
\pgfplotstableread{100t1b3i1_5.dat}\blurat

\addplot +[quartile] table[x=photon_nums,y=q1] \raw;
\addplot +[med] table[x=photon_nums,y=med] \raw;           \addlegendentry{No Correction}
\addplot +[quartile] table[x=photon_nums,y=q3] \raw;

\pgfplotsset{cycle list={{blue!60!black}}}

\addplot +[quartile] table[x=photon_nums,y=q1] \blura;
\addplot +[med,dashed] table[x=photon_nums,y=med] \blura; \addlegendentry{DGC}
\addplot +[quartile] table[x=photon_nums,y=q3] \blura;

\pgfplotsset{cycle list={{black}}}

\addplot +[med] table[x=photon_nums,y=med] \rawt; \addlegendentry{AWTF}

\def\ymax{4e-6}
\draw [dotted,black](axis cs:27542,1e-8)-- (axis cs:27542,\ymax)
node[pos=0,rotate=90,anchor=north west] {$m=14$}
node[pos=1,rotate=90,anchor=north east] {$m=16$};
\draw [dotted,black](axis cs:4365,1e-8)-- (axis cs:4365,\ymax)
node[pos=0,rotate=90,anchor=north west] {$m=16$}
node[pos=1,rotate=90,anchor=north east] {$m=18$};
\draw [dotted,black](axis cs:692,1e-8)-- (axis cs:692,\ymax)
node[pos=0,rotate=90,anchor=north west] {$m=18$}
node[pos=1,rotate=90,anchor=north east] {$m=20$};
\draw [dotted,black](axis cs:109.6,1e-8)-- (axis cs:109.6,\ymax)
node[pos=0,rotate=90,anchor=north west] {$m=20$}
node[pos=1,rotate=90,anchor=north east] {$m=22$};

\end{loglogaxis}
\end{tikzpicture}
\end{figure*}

There is a clear advantage, shown in \fref{fig:strats}, of applying some correction to the measured intensity images before the reconstruction occurs. DGC is particularly fast to  execute, either on a GPU or within a standard PC. The good performance of DGC is perhaps because, unlike using spatial binning, the location information of the photons counted is preserved, albeit slightly `smeared'. This allows the nlCWFS to more accurately determine the phase, especially the low order modes, since these are more spatially spread regardless of light level.  AWTF offers a slight improvement over DGC, although it will execute more slowly given the  computational complexity of the technique. 

All of the techniques investigated offer improvements at low photon counts at the expense of accuracy at high photon-counts. This doesn't present a problem however, as these corrections can be activated at will, and are not `built in' to the algorithm itself, merely added at the start. A good estimate of the crossover points at which each approach becomes best-suited is all that is needed to ensure optimum performance.


\begin{figure*}[htpb]
\caption{Simulated low light performance of CWFS using algorithm IO4 preprocessing the input planes with DGC3. Vertical lines show the I-band magnitude required for a given photon count with the CWFS running at $10\hz$ for $D=4.2\m$ (lower label), and $D=10.5\m$ (upper label). The curves show the performance for different Lucky Imaging selection percentages.}
\label{fig:qnts:dgc3}
\begin{tikzpicture}[node distance = 0.05cm]
\begin{loglogaxis}[
    width=2\columnwidth,
    height=10cm,
    ylabel={RMS optical path difference $/\m$},
    xlabel={Photons detected},
    xmin=30, xmax=1000000,
    ymin=1e-8, ymax=2e-6,
    minmax/.style={dotted,mark=none,forget plot},
    quartile/.style={dotted,mark=none,forget plot},
    med/.style={solid,mark=none}
]

\pgfplotstableread{100qt0b3i1_5.dat}\blurq

\pgfplotscreateplotcyclelist{linestyles}{solid,dashed,dotted}

\pgfplotsset{cycle list name=linestyles}

\addplot  table[x=photon_nums,y=75pc] \blurq; \addlegendentry{75\%}
\addplot  table[x=photon_nums,y=50pc] \blurq; \addlegendentry{50\%}
\addplot  table[x=photon_nums,y=25pc] \blurq; \addlegendentry{25\%}
\addplot  table[x=photon_nums,y=10pc] \blurq; \addlegendentry{10\%}
\addplot  table[x=photon_nums,y=5pc] \blurq; \addlegendentry{5\%}
\addplot  table[x=photon_nums,y=1pc] \blurq; \addlegendentry{1\%}

\def\ymax{2e-6}
\draw [dotted,black](axis cs:27542,1e-8)-- (axis cs:27542,\ymax)
node[pos=0,rotate=90,anchor=north west] {$m=14$}
node[pos=1,rotate=90,anchor=north east] {$m=16$};
\draw [dotted,black](axis cs:4365,1e-8)-- (axis cs:4365,\ymax)
node[pos=0,rotate=90,anchor=north west] {$m=16$}
node[pos=1,rotate=90,anchor=north east] {$m=18$};
\draw [dotted,black](axis cs:692,1e-8)-- (axis cs:692,\ymax)
node[pos=0,rotate=90,anchor=north west] {$m=18$}
node[pos=1,rotate=90,anchor=north east] {$m=20$};
\draw [dotted,black](axis cs:109.6,1e-8)-- (axis cs:109.6,\ymax)
node[pos=0,rotate=90,anchor=north west] {$m=20$}
node[pos=1,rotate=90,anchor=north east] {$m=22$};

\end{loglogaxis}
\end{tikzpicture}
\end{figure*}
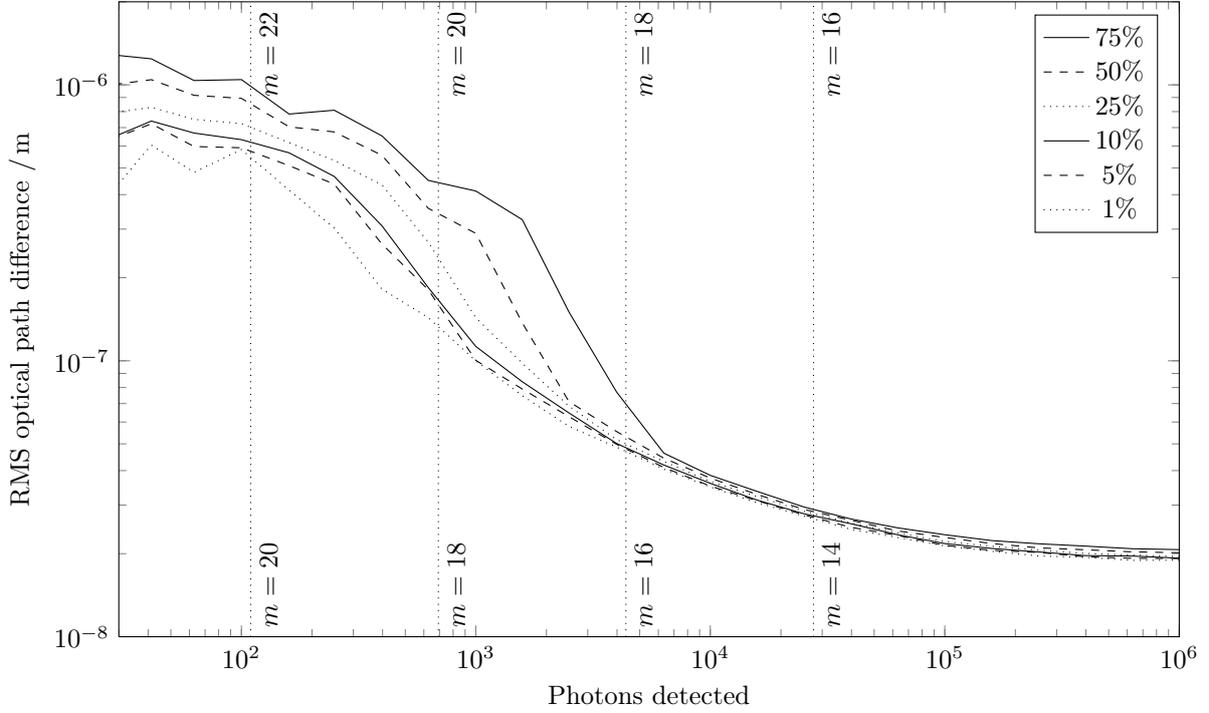

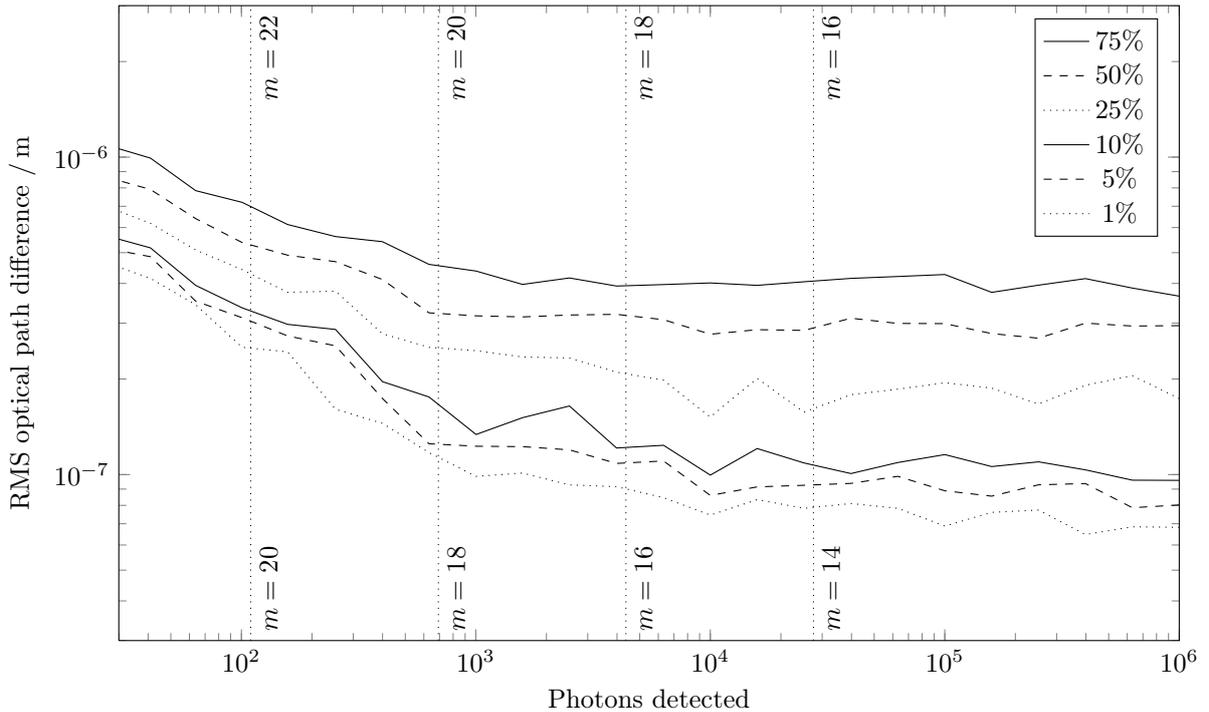
\begin{figure*}[htpb]
\caption{Simulated low light performance of CWFS using algorithm IO4 preprocessing the input planes with DGC8. Vertical lines show the I-band magnitude required for a given photon count with the CWFS running at $10\hz$ for $D=4.2\m$ (lower label), and $D=10.5\m$ (upper label).}
\label{fig:qnts:dgc8}
\begin{tikzpicture}[node distance = 0.05cm]
\begin{loglogaxis}[
    width=2\columnwidth,
    height=10cm,
    ylabel={RMS optical path difference $/\m$},
    xlabel={Photons detected},
    xmin=30, xmax=1000000,
    ymin=3e-8, ymax=3e-6,
    minmax/.style={dotted,mark=none,forget plot},
    quartile/.style={dotted,mark=none,forget plot},
    med/.style={solid,mark=none}
]

\pgfplotstableread{100qt0b8i4.dat}\blurq
\pgfplotscreateplotcyclelist{linestyles}{solid,dashed,dotted}

\pgfplotsset{cycle list name=linestyles}

\addplot  table[x=photon_nums,y=75pc] \blurq; \addlegendentry{75\%}
\addplot  table[x=photon_nums,y=50pc] \blurq; \addlegendentry{50\%}
\addplot  table[x=photon_nums,y=25pc] \blurq; \addlegendentry{25\%}
\addplot  table[x=photon_nums,y=10pc] \blurq; \addlegendentry{10\%}
\addplot  table[x=photon_nums,y=5pc] \blurq; \addlegendentry{5\%}
\addplot  table[x=photon_nums,y=1pc] \blurq; \addlegendentry{1\%}

\def\ymax{3e-6}
\def\ymin{3e-8}
\draw [dotted,black](axis cs:27542,\ymin)-- (axis cs:27542,\ymax)
node[pos=0,rotate=90,anchor=north west] {$m=14$}
node[pos=1,rotate=90,anchor=north east] {$m=16$};
\draw [dotted,black](axis cs:4365,\ymin)-- (axis cs:4365,\ymax)
node[pos=0,rotate=90,anchor=north west] {$m=16$}
node[pos=1,rotate=90,anchor=north east] {$m=18$};
\draw [dotted,black](axis cs:692,\ymin)-- (axis cs:692,\ymax)
node[pos=0,rotate=90,anchor=north west] {$m=18$}
node[pos=1,rotate=90,anchor=north east] {$m=20$};
\draw [dotted,black](axis cs:109.6,\ymin)-- (axis cs:109.6,\ymax)
node[pos=0,rotate=90,anchor=north west] {$m=20$}
node[pos=1,rotate=90,anchor=north east] {$m=22$};

\end{loglogaxis}
\end{tikzpicture}
\end{figure*}

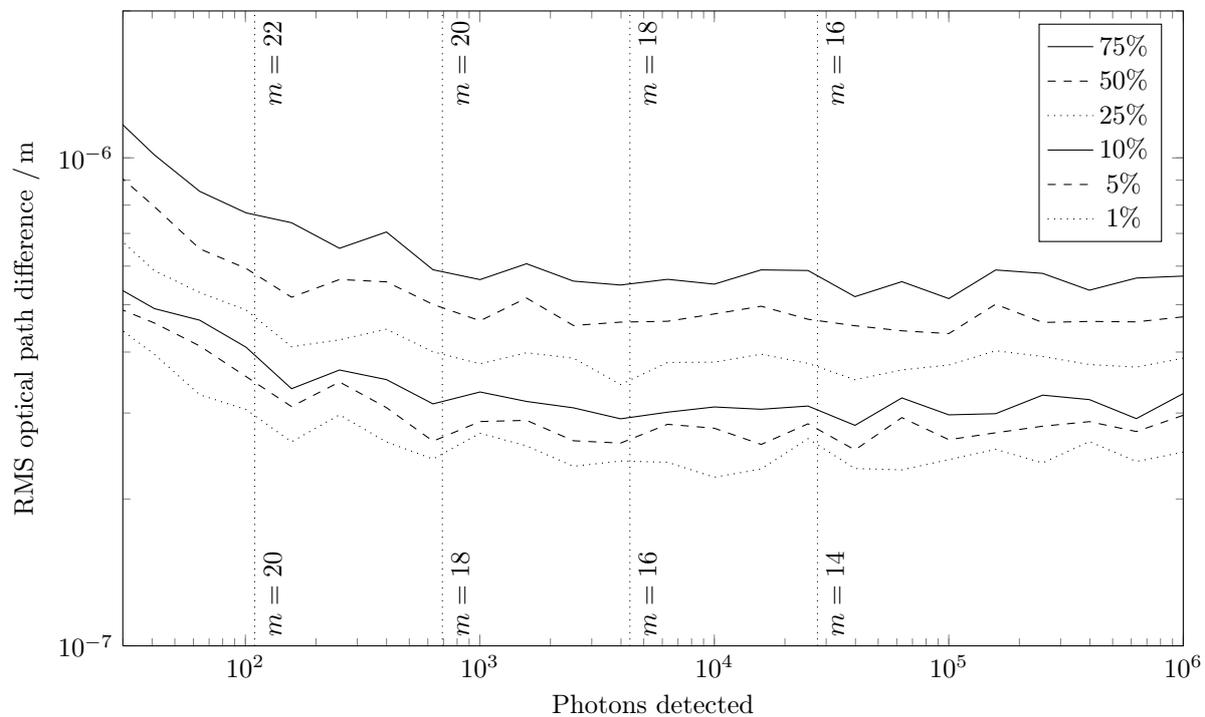
\begin{figure*}[htpb]
\caption{Simulated low light performance of CWFS using algorithm IO4 preprocessing the input planes with DGC16. Vertical lines show the I-band magnitude required for a given photon count with the CWFS running at $10\hz$ for $D=4.2\m$ (lower label), and $D=10.5\m$ (upper label).}
\label{fig:qnts:dgc16}
\begin{tikzpicture}[node distance = 0.05cm]
\begin{loglogaxis}[
    width=2\columnwidth,
    height=10cm,
    ylabel={RMS optical path difference $/\m$},
    xlabel={Photons detected},
    xmin=30, xmax=1000000,
    ymin=1e-7, ymax=2e-6,
    minmax/.style={dotted,mark=none,forget plot},
    quartile/.style={dotted,mark=none,forget plot},
    med/.style={solid,mark=none}
]

\pgfplotstableread{100qt0b16i8.dat}\blurq

\pgfplotscreateplotcyclelist{linestyles}{solid,dashed,dotted}

\pgfplotsset{cycle list name=linestyles}

\addplot  table[x=photon_nums,y=75pc] \blurq; \addlegendentry{75\%}
\addplot  table[x=photon_nums,y=50pc] \blurq; \addlegendentry{50\%}
\addplot  table[x=photon_nums,y=25pc] \blurq; \addlegendentry{25\%}
\addplot  table[x=photon_nums,y=10pc] \blurq; \addlegendentry{10\%}
\addplot  table[x=photon_nums,y=5pc] \blurq; \addlegendentry{5\%}
\addplot  table[x=photon_nums,y=1pc] \blurq; \addlegendentry{1\%}

\def\ymax{2e-6}
\def\ymin{1e-7}
\draw [dotted,black](axis cs:27542,\ymin)-- (axis cs:27542,\ymax)
node[pos=0,rotate=90,anchor=north west] {$m=14$}
node[pos=1,rotate=90,anchor=north east] {$m=16$};
\draw [dotted,black](axis cs:4365,\ymin)-- (axis cs:4365,\ymax)
node[pos=0,rotate=90,anchor=north west] {$m=16$}
node[pos=1,rotate=90,anchor=north east] {$m=18$};
\draw [dotted,black](axis cs:692,\ymin)-- (axis cs:692,\ymax)
node[pos=0,rotate=90,anchor=north west] {$m=18$}
node[pos=1,rotate=90,anchor=north east] {$m=20$};
\draw [dotted,black](axis cs:109.6,\ymin)-- (axis cs:109.6,\ymax)
node[pos=0,rotate=90,anchor=north west] {$m=20$}
node[pos=1,rotate=90,anchor=north east] {$m=22$};

\end{loglogaxis}
\end{tikzpicture}
\end{figure*}

As can be seen in Figures \ref{fig:qnts:dgc3}--\ref{fig:qnts:dgc16}, the level of blurring used places a lower bound on the residual OPD. The data suggest that smaller amounts of blurring allow better correction at higher photon-counts; higher amounts of blurring limit the level of correction attainable, but increase the likelihood of achieving a similar level of correction at low photon-counts. Using DGC with 3px outer blurring and 1.5px inner (henceforth, DGC$n$ refers to outer spacing of $n$ px and inner spacing of $\frac{n}{2}$ px), 10\% of inputs were corrected to around $100\nm$ residual RMS, with similar performance for DGC8 (\fref{fig:qnts:dgc8}), but DGC16 (\fref{fig:qnts:dgc16}) gave poorer results at this level. Comparing the size of the gaussian blur used to the speckle size gives an insight into this. At $\pm z_2$, we expect a speckle size of $\frac{\lambda z}{D}=44.6\cm \simeq 9\px$. At $\pm z_1$ we expect a speckle size of $17.9\cm\simeq 4\px$. These values closely match those used for DGC8. A possible reason for DGC8 allowing good corrections at low photon counts is its rejection (by blurring) of features smaller than the speckle size.

\begin{figure*}[htpb]
\caption{Simulated low light performance of CWFS using algorithm IO4 preprocessing the input planes with GV. Vertical lines show the I-band magnitude required for a given photon count with the CWFS running at $10\hz$ for $D=4.2\m$ (lower label), and $D=10.5\m$ (upper label).}
\label{fig:qnts:v4b4}
\centering
\begin{tikzpicture}[node distance = 0.05cm]
\begin{loglogaxis}[
    width=1.5\columnwidth,
    height=10cm,
    ylabel={RMS optical path difference $/\m$},
    xlabel={Photons detected},
    xmin=30, xmax=1000,
    ymin=0.2e-7, ymax=2e-6,
    minmax/.style={dotted,mark=none,forget plot},
    quartile/.style={dotted,mark=none,forget plot},
    med/.style={solid,mark=none},
    every axis legend/.append style={
    	at={(0.02,0.02)},
	anchor=south west,
    }
]

\pgfplotstableread{100v4b4.dat}\blurq

\pgfplotscreateplotcyclelist{linestyles}{solid,dashed,dotted}

\pgfplotsset{cycle list name=linestyles}

\addplot  table[x=photon_nums,y=75pc] \blurq; \addlegendentry{75\%}
\addplot  table[x=photon_nums,y=50pc] \blurq; \addlegendentry{50\%}
\addplot  table[x=photon_nums,y=25pc] \blurq; \addlegendentry{25\%}
\addplot  table[x=photon_nums,y=10pc] \blurq; \addlegendentry{10\%}
\addplot  table[x=photon_nums,y=5pc] \blurq; \addlegendentry{5\%}
\addplot  table[x=photon_nums,y=1pc] \blurq; \addlegendentry{1\%}

\def\ymax{2e-6}
\def\ymin{2e-8}
\draw [dotted,black](axis cs:692,\ymin)-- (axis cs:692,\ymax)
node[pos=0,rotate=90,anchor=north west] {$m=18$}
node[pos=1,rotate=90,anchor=north east] {$m=20$};
\draw [dotted,black](axis cs:109.6,\ymin)-- (axis cs:109.6,\ymax)
node[pos=0,rotate=90,anchor=north west] {$m=20$}
node[pos=1,rotate=90,anchor=north east] {$m=22$};

\end{loglogaxis}
\end{tikzpicture}
\end{figure*}
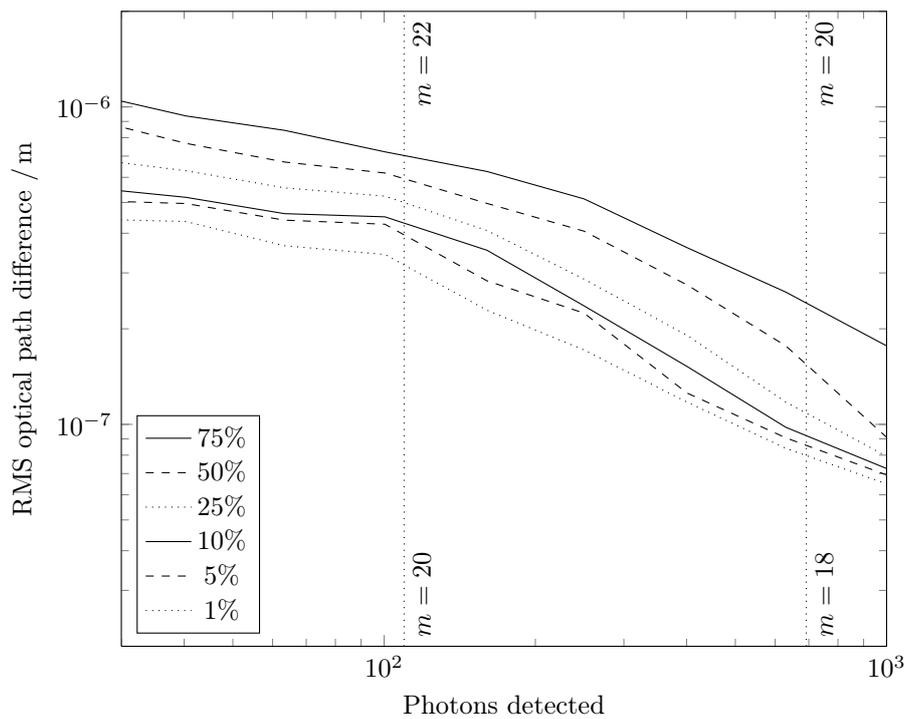

Preprocessing the input with GV (\fref{fig:qnts:v4b4}) gives improved performance over DGC8 with 300--1000 photons. In the PSF shown in \fref{fig:psf_gv}, the first Airy disc is just distinguishable, although the halo is of comparable size to the uncorrected image, if somewhat fainter. This is consistent with removal of the lower order errors, which is precisely what is required for Lucky Imaging. 

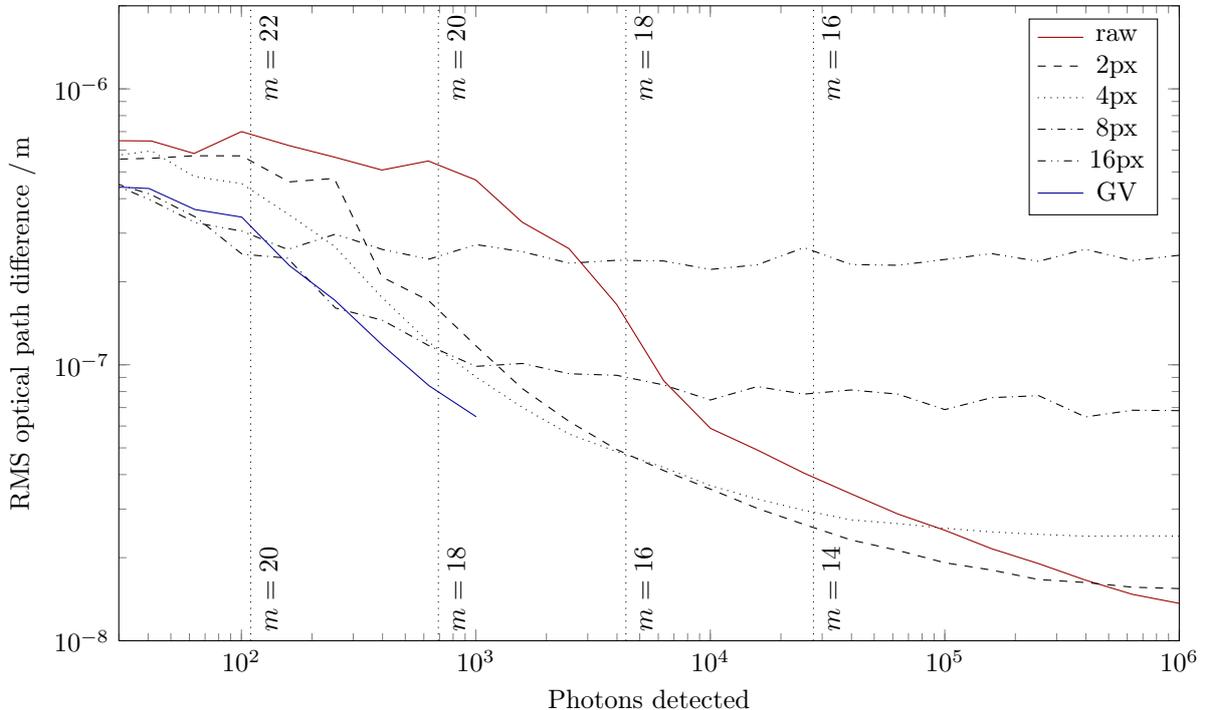
\begin{figure*}[htpb]
\caption{DGC performance: The 1\% quantile is shown for different amounts of blurring and for GV. For DGC amount for the inner planes is half that which is shown for the outer planes. GV is applied only at photon counts of 1000 and below. Vertical lines show the I-band magnitude required for a given photon count with the CWFS running at $10\hz$ for $D=4.2\m$ (lower label), and $D=10.5\m$ (upper label).}
\label{fig:qnts:dgc_all}
\begin{tikzpicture}[node distance = 0.05cm]
\begin{loglogaxis}[
    width=2\columnwidth,
    height=10cm,
    ylabel={RMS optical path difference $/\m$},
    xlabel={Photons detected},
    xmin=30, xmax=1000000,
    ymin=1e-8, ymax=2e-6,
    minmax/.style={dotted,mark=none,forget plot},
    quartile/.style={dotted,mark=none,forget plot},
    med/.style={solid,mark=none}
]

\pgfplotstableread{100qt0b2i1.dat}\blurb
\pgfplotstableread{100qt0b4i2.dat}\blurd
\pgfplotstableread{100qt0b8i4.dat}\blure
\pgfplotstableread{100qt0b16i8.dat}\blurf
\pgfplotstableread{100v4b4.dat}\vor


\pgfplotstableread{100t0b0.dat}\raw

\pgfplotsset{cycle list name=linestyles*}

\addplot +[red!60!black] table[x=photon_nums,y=min] \raw; \addlegendentry{raw}
\addplot  table[x=photon_nums,y=1pc] \blurb; \addlegendentry{2px}
\addplot  table[x=photon_nums,y=1pc] \blurd; \addlegendentry{4px}
\addplot  table[x=photon_nums,y=1pc] \blure; \addlegendentry{8px}
\addplot  table[x=photon_nums,y=1pc] \blurf; \addlegendentry{16px}

\addplot [blue!60!black] table[x=photon_nums,y=1pc] \vor; \addlegendentry{GV}


\def\ymax{2e-6}
\draw [dotted,black](axis cs:27542,1e-8)-- (axis cs:27542,\ymax)
node[pos=0,rotate=90,anchor=north west] {$m=14$}
node[pos=1,rotate=90,anchor=north east] {$m=16$};
\draw [dotted,black](axis cs:4365,1e-8)-- (axis cs:4365,\ymax)
node[pos=0,rotate=90,anchor=north west] {$m=16$}
node[pos=1,rotate=90,anchor=north east] {$m=18$};
\draw [dotted,black](axis cs:692,1e-8)-- (axis cs:692,\ymax)
node[pos=0,rotate=90,anchor=north west] {$m=18$}
node[pos=1,rotate=90,anchor=north east] {$m=20$};
\draw [dotted,black](axis cs:109.6,1e-8)-- (axis cs:109.6,\ymax)
node[pos=0,rotate=90,anchor=north west] {$m=20$}
node[pos=1,rotate=90,anchor=north east] {$m=22$};

\end{loglogaxis}
\end{tikzpicture}
\end{figure*}

From \fref{fig:qnts:dgc_all}, a correction strategy can be derived, which is shown in \tref{tab:dgcstrat}.

\begin{table}[h]
\caption{A correction strategy using DGC}
\label{tab:dgcstrat}
\centering
\begin{tabular}{| r @{  }  r | c |}
\hline
\multicolumn{2}{|c|}{\# photons} & \T\B blur amount (px) \\
\hline
\T> & $4\cdot10^5$ & 0\\ 
> & $4\cdot10^3$ & 2\\ 
> & $6\cdot10^2$ & 4\\ 
\multicolumn{2}{|c|}{otherwise}& \B 8\\ \hline
\end{tabular}
\end{table}

\subsection{Comparing the simulation to expected conditions}

With a better model for generating temporally-related phase-screens, the simulations could have been made more realistic. As each pupil phase was randomly generated, there was no relationship between successive phases. In practice, however, successive pupil phases will be related. With a constant wind-speed, the large-scale structure of the turbulence, and of the phase error, will move across the aperture at a constant speed. Identifying the speed and direction of this motion would allow a reasonable correction of the low-order phase errors, thus reducing the RMS OPD of the light entering the CWFS. In fact, a whole host of finely-tuned motion-estimation algorithms do already exist, employed for the compression of video images.  Typical median wind speeds on good astronomical sites are in the region of 8-10 m/s.

This lack of prior phase estimate in the simulation decreases the performance of the algorithm. However, the selected value of $r_0$ gave an RMS OPD similar to that which it is estimated would be encountered in closed-loop AO.

\Citeauthor{AObook} gives $m=14.6+8 \log(\lambda_{\um})$ as an estimate of the limiting magnitude for AO using a SHWFS (\citeyear{AObook}) --- a compromise between reducing the number of sub-apertures (to get more light per subaperture) and the seeing limit that this imposes due to the subaperture-size being greater than $r_0$. This gives a value of $m=13.6$ for $\lambda=750\nm$. The performance that could be delivered by the present technique will allow the use of guide-stars up to 5 magnitudes fainter than this.

\section{Experimental comparison of algorithms}
\label{sec:exp}

The focus of this section is the relative performance of the reconstruction algorithms, so only a few fixed photon-count values have been used. All algorithms perform well in ample light, which is not the operating region of LI, so less focus has been placed on exploring this region. Where performance is of interest is in the region where the RMSE is of order a few radians, as this is approximately the working range for LI. Three light levels have been investigated --- 100, 1,000, and 10,000 photons. The pre-processing used for each scenario, shown in \tref{tab:expstrat}, is based on the strategy outlined in \sref{sec:lowlight:compstrat} (\tref{tab:dgcstrat}). The statistics shown are based on 100 independent runs, as in \sref{sec:lowlight}.

\begin{table}[h!]
\caption{Pre-processing used for comparing parameter values and algorithms}
\label{tab:expstrat}
\centering
\begin{tabular}{| c | c | c |}
\hline
\multicolumn{1}{| c |}{\multirow{2}{*}{\# photons}}&\multicolumn{2}{ c |}{ \T \B blur amount (px)}\\ \cline{2-3}
\multicolumn{1}{| c |}{}& \T outer &\B inner\\ \hline
100		& 8 & 4 \\
1,000	& 4 & 2 \\
10,000	& 2 &\B 1 \\
\hline
\end{tabular}
\end{table}

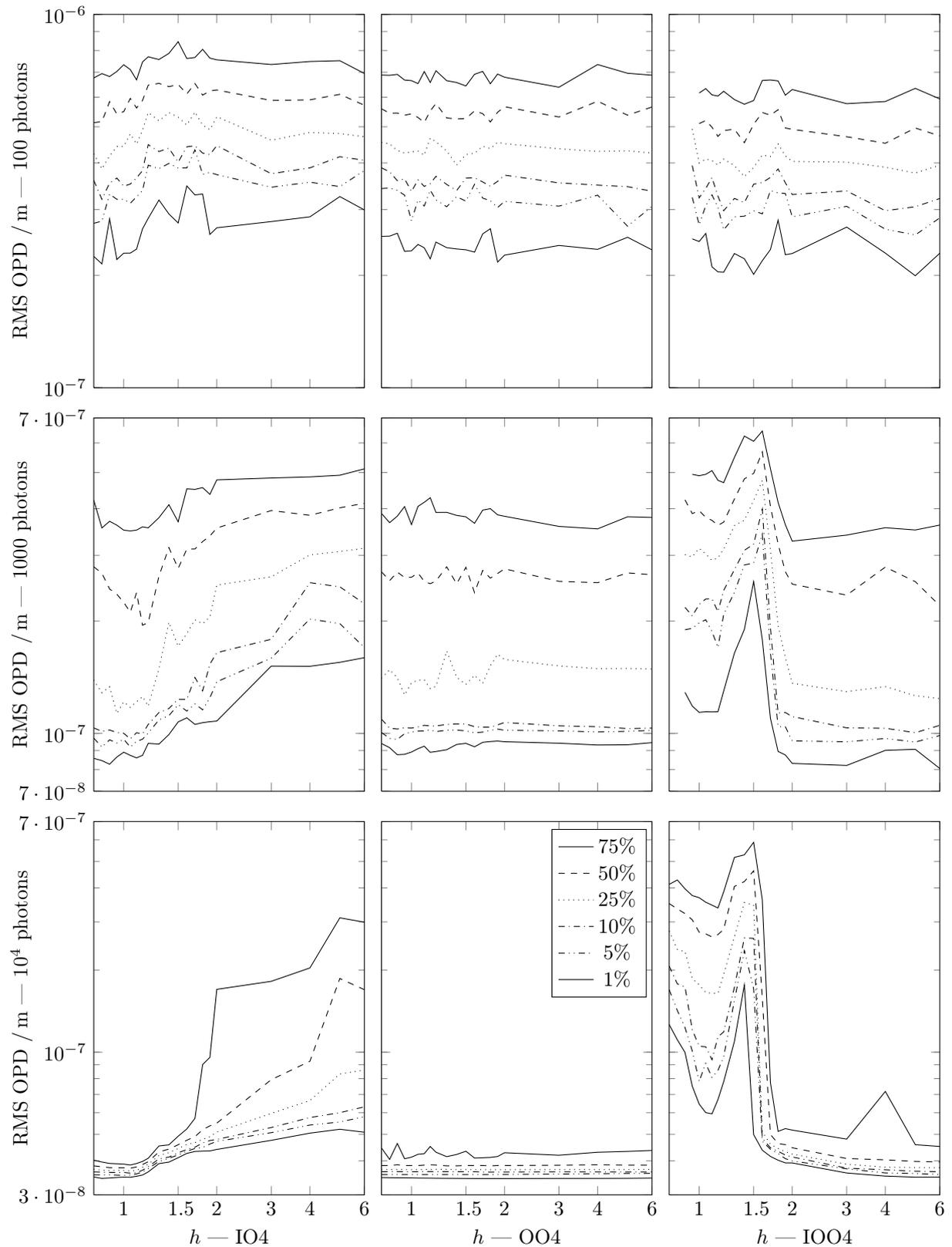
\begin{figure*}
\caption{Performance of the algorithms at different light levels. The logarithmic horizontal scale shows the parameter $\mathit{h}$ used, whilst the vertical axis shows the RMS OPD after correction (n.b. different scale for each row). The absence of points at the 100- and 1000-photon levels in the IOO4 algorithm is due to the failure of the algorithm to converge to within the tolerance of the error unwrapping algorithm in any of the samples. The legend in the lower central panel applies to all of the graphs.}
\label{fig:hcomp}
 \centering
\begin{tikzpicture}

\pgfplotsset{
every axis/.append style= {
	xmin=0.8,
	xmax=6,
	ymin=1e-7,
	ymax=1e-6,
	ylabel absolute,
	ylabel shift=-1em,
	xtick={1,1.5,2,3,4,6},
	xticklabels={1,1.5,2,3,4,6},
	width=6.2cm,
	height=8cm
	},	
xticklabel style={
	text width=1em,
	align=center
	}
}


\begin{loglogaxis}[
	name=p11,
	ylabel={RMS OPD $/\m$ --- 100 photons},
	xticklabels={}
]
\pgfplotsset{cycle list name=linestyles*}
\foreach \x in {6,5,...,1}{
\addplot table[header=false, x index=0,y index=\x] {hq_a1_p1.dat};
}
\end{loglogaxis}

\begin{loglogaxis}[
	name=p21,
	at=(p11.right of east),
	anchor=west,
	yticklabels={}
	xlabel={},
	xticklabels={}
]
\pgfplotsset{cycle list name=linestyles*}
\foreach \x in {6,5,...,1}{
\addplot table[header=false, x index=0,y index=\x] {hq_a2_p1.dat};
}
\end{loglogaxis}

\begin{loglogaxis}[
	name=p31,
	at=(p21.right of east),
	anchor=west,
	yticklabels={}
	xlabel={},
	xticklabels={}
]
\pgfplotsset{cycle list name=linestyles*}
\foreach \x in {6,5,...,1}{
\addplot table[header=false, x index=0,y index=\x] {hq_a3_p1.dat};
}

\end{loglogaxis}


\begin{loglogaxis}[
	name=p12,
	at=(p11.below south),
	anchor=above north,
	ylabel={RMS OPD $/\m$ --- 1000 photons},
	xticklabels={},
	ymin=7e-8,
	ymax=7e-7,
	minor ytick={7e-8,8e-8,9e-8,1e-7,2e-7,3e-7,4e-7,5e-7,6e-7,7e-7},
	ytick={7e-8,1e-7,7e-7},
	yticklabels={$7\cdot10^{-8}$,$10^{-7}$,$7\cdot10^{-7}$,}
]
\pgfplotsset{cycle list name=linestyles*}
\foreach \x in {6,5,...,1}{
\addplot table[header=false, x index=0,y index=\x] {hq_a1_p2.dat};
}
\end{loglogaxis}

\begin{loglogaxis}[
	name=p22,
	at=(p12.right of east),
	anchor=west,
	yticklabels={},
	xticklabels={},
	ymin=7e-8,
	ymax=7e-7,
	minor ytick={7e-8,8e-8,9e-8,1e-7,2e-7,3e-7,4e-7,5e-7,6e-7,7e-7},
	ytick={7e-8,1e-7,7e-7}
]
\pgfplotsset{cycle list name=linestyles*}
\foreach \x in {6,5,...,1}{
\addplot table[header=false, x index=0,y index=\x] {hq_a2_p2.dat};
}
\end{loglogaxis}

\begin{loglogaxis}[
	name=p32,
	at=(p22.right of east),
	anchor=west,
	yticklabels={},
	xticklabels={},
	ymin=7e-8,
	ymax=7e-7,
	minor ytick={7e-8,8e-8,9e-8,1e-7,2e-7,3e-7,4e-7,5e-7,6e-7,7e-7},
	ytick={7e-8,1e-7,7e-7}
]
\pgfplotsset{cycle list name=linestyles*}
\foreach \x in {6,5,...,1}{
\addplot table[header=false, x index=0,y index=\x] {hq_a3_p2.dat};
}
\end{loglogaxis}


\begin{loglogaxis}[
	name=p13,
	at=(p12.below south),
	anchor=above north,
	xlabel={$h$ --- IO4},
	ylabel={RMS OPD $/\m$ --- $10^4$ photons},
	ymin=3e-8,
	ymax=7e-7,
	minor ytick={3e-8,4e-8,5e-8,6e-8,7e-8,8e-8,9e-8,1e-7,2e-7,3e-7,4e-7,5e-7,6e-7,7e-7},
	ytick={3e-8,1e-7,7e-7},
	yticklabels={$3\cdot10^{-8}$,$10^{-7}$,$7\cdot10^{-7}$,}
]
\pgfplotsset{cycle list name=linestyles*}
\foreach \x in {6,5,...,1}{
\addplot table[header=false, x index=0,y index=\x] {hq_a1_p3.dat};
}
\end{loglogaxis}

\begin{loglogaxis}[
	name=p23,
	at=(p13.right of east),
	anchor=west,
	xlabel={$h$ --- OO4},
	yticklabels={},
	ymin=3e-8,
	ymax=7e-7,
	minor ytick={3e-8,4e-8,5e-8,6e-8,7e-8,8e-8,9e-8,1e-7,2e-7,3e-7,4e-7,5e-7,6e-7,7e-7},
	ytick={3e-8,1e-7,7e-7}
]
\pgfplotsset{cycle list name=linestyles*}
\foreach \x in {6,5,...,1}{
\addplot table[header=false, x index=0,y index=\x] {hq_a2_p3.dat};
}
\legend{75\%,50\%,25\%,10\%,5\%,1\%}
\end{loglogaxis}

\begin{loglogaxis}[
	name=p33,
	at=(p23.right of east),
	anchor=west,
	xlabel={$h$ --- IOO4},
	yticklabels={},
	ymin=3e-8,
	ymax=7e-7,
	minor ytick={3e-8,4e-8,5e-8,6e-8,7e-8,8e-8,9e-8,1e-7,2e-7,3e-7,4e-7,5e-7,6e-7,7e-7},
	ytick={3e-8,1e-7,7e-7}
]
\pgfplotsset{cycle list name=linestyles*}
\foreach \x in {6,5,...,1}{
\addplot table[header=false, x index=0,y index=\x] {hq_a3_p3.dat};
}
\end{loglogaxis}

\end{tikzpicture}
\end{figure*}

With $10^4$ photons, IO4 has marginally better performance than OO4 or IOO4, with even the 75th percentile having very low error of $40\nm$ RMS. At this light level, IO4 performs best for $h\sim1$, whilst OO4 seems agnostic to step size, but has a slightly increased error in the higher percentiles. This level of performance would be important for ordinary imaging, but is slightly less important for LI, which provides no real advantage in this region. The curious dependence of IOO4 on $h$ seems to occur at all light levels. The data suggest the idea that two operating regions exist, one either side of this local error maximum. How these regions differ, and why it should be that the algorithm converges so poorly with intermediate $h$ values is not immediately apparent. It is possible that this behaviour may be related to the arbitrary phase difference discussed in \sref{sec:alg:arbphase}. The wide variation in performance of these algorithms suggests that a more exhaustive search of the parameter space could yield algorithms with better performance. 

When the number of photons is reduced to 1000, the IO4 algorithm has the highest proportion of output residuals below $100\nm$ RMS, with 10\% of outputs falling below this value. Even at 25\% selection for LI, good images would likely be obtained. OO4 and IOO4 also perform well at this light-level, each behaving as at higher light levels, but with higher error, and a lower proportion of outputs near the minimum error.

Upon further reduction of the number of photons to 100, algorithm IOO4 with $h=5$ outperforms the other two algorithms for the parameter values tested, with the minimum of each percentile lower than the corresponding minimum for IO4 or OO4. In designing an AO system, the data in \fref{fig:hcomp} enable the selection of algorithm and parameters best matched to the conditions and goals of use.

\section{Discussion}
\label{sec:discuss}

Reconstruction of wavefront phase from photon-limited defocus images is an unusual task, and there does not appear to have been a significant amount of research into this very particular application. A statistically driven (e.g. Bayesian) reconstruction strategy may improve further on the results obtained here, which would in turn allow for improved algorithm performance at low photon-counts.

The performance at low light levels here suggests that a nlCWFS is a good solution to shortcomings of the SHWFS particularly at low light levels and low-order phase errors. Even with an input phase error of $5\rad$ RMS, nlCWFS + DGC is able to reduce the phase error into LI territory with as few as 1000 photons. If this performance is achieved on the Gran Telescopio Canarias (GTC), natural guide stars with I-band magnitude 20 could be used. With a sufficiently fast implementation of GV preprocessing, the limits could extend even further into the low light levels, as this technique delivers good results even with only 400 photons.

Although the work presented in \sref{sec:exp} has shown an investigation of several algorithms, it represents only a small subset of possible algorithms --- a few straight lines through the multidimensional space of $a_i$ and $b_i$. In fact, the search here has been limited to the 4-brane passing through $a_0=1, b_0=1, a_1=1, \textrm{and } b_1=1$. An optimisation approach could be used to find values of $a_{i}, b_{i}$ that perform best at a given photon level. This would require considerable computing time, as each parameter optimised adds a dimension to the search space. As an example, finding optimal parameters for $a_{i}, b_{i}, i<q$ to the nearest $d$, and constraining $|a_i|,|b_i|<c$ would require comparing the performance of $\left(2\frac{c}{d}+1\right)^{(2q-1)}$ algorithms. Constraining the search space by applying heuristics to determine which algorithms might be `forward-stepping' rather than `backward-stepping' could reduce the computational cost.

At very low light-levels, an `adaptive' algorithm may outperform those investigated here. Such an algorithm would, instead of rigidly imposing amplitude constraints, refine the estimate of the amplitude constraints, taking into account the fact that the input data is subject to Poisson statistics and is therefore not a fair representation of the true intensities at the measurement planes. 

An atmospheric model capable of using previous phase aberrations to infer the motions of turbulent structures across the aperture would be able to provide a much better initial estimate for the phase, and therefore the algorithm would start much closer to the desired solution. There is, however, a fundamental limit to what information we can deduce from a given number of photons. In order to drive an $n$-element deformable mirror, we need phase information about each of the $n$ elements, which, in our 4-plane detector, comes from measurement of $\sim O(4n)$ photons.

It is hoped that further refinement of these algorithms together with improvements in the pre-processing of photon-limited defocus images will allow an extension of AO capabilities down to the lowest light-levels. This would make Lucky Imaging a powerful technique for ground-based imaging of faint objects at high angular resolution.

The authors would like to acknowledge helpful discussions and advice from J. R. Fienup.


\label{sec:ref}

\printbibliography

\vfill
\eject 

\end{document}